\documentclass[11pt]{article}
\usepackage{geometry}                
\usepackage[parfill]{parskip}    
\usepackage{graphicx}
\usepackage{amsmath, amsthm, amsfonts}
\usepackage{amssymb}

\usepackage{color}
\usepackage[colorlinks]{hyperref}
\hypersetup{linkcolor=blue,%
citecolor=red,%
urlcolor=cyan}
\usepackage{url} 

\usepackage{epstopdf}
\title{\bf  A dynamical approach to relativity: Desynchronization of clocks in rigid acceleration and gravitational redshift}

\author { \.{I}nan\c{c} \c{S}ahin $^a$\footnote{inancsahin@ankara.edu.tr}, 
\date{}                        

\begin{document}

\maketitle

\begin{abstract}
In this paper, inspired by quantum field theory, or more specifically QED, we propose a dynamical model for relativity. By adopting the approach provided by this dynamical model, we provide a dynamical explanation for relativistic phenomena such as length contraction, time dilation and clock desynchronization. The result of the dynamical approach for clock desynchronization is especially interesting. Our dynamical approach exhausts the freedom for different clock synchronizations on frames accelerated from an initial reference frame, which is allowed by Einstein's principle approach.
Furthermore, our dynamical approach arrives at the interesting conclusion that Einstein's equivalence principle and gravitational redshift are necessary to explain special relativistic clock desynchronization. This result shows that special relativity by itself is an incomplete theory. If our dynamical model assumptions are assumed to be valid at a fundamental level, it implies some modifications to quantum field theory and general relativity via the equivalence principle. These modifications have been discussed.
 
\end{abstract}



\section{Introduction}
\label{intro}
Rigid acceleration is a type of acceleration in which the proper lengths in the accelerating frame remain the same. 
Therefore, in such an acceleration, the rigidity of a body is maintained, it does not deform or change shape in its rest frame due to the acceleration forces. In this paper we will examine the behavior of rigidly accelerating rods and clocks. Unlike Einstein's original principle approach to relativity \cite{Einstein1919,Bell,Brown2005}, we adopt a kind of dynamical approach and derive length contraction, time dilation and clock desynchronization as a result of some underlying processes. As it was pointed out by Bell \cite{Bell}, Einstein's approach to relativity differs from that of Lorentz and others in two major ways. There is a difference of philosophy, and a difference of style. The difference of style is that Einstein started from the assumption that laws would appear the same to all inertial observers. But he did not infer observers' experiences from underlying physical processes. Einstein did not create a theory of clocks and duration from first principles. The properties of clocks were not deduced from the inner structure of the theory, but were simply required to accord with the relativity principle \cite{Barbour}. Einstein's approach  permits a very concise and elegant formulation of the theory, however his approach leaves some details unanswered. 
For instance, consider spatially seperated clocks with a proper distance $\ell$ synchronized in an inertial frame $S_0$. Then, they are accelerated into another inertial frame say $S_1$ which has a relative speed $v$ with respect to $S_0$. From Lorentz transformations we may expect that according to an observer in $S_0$ the trailing clock will shift forward $\Delta t=\beta\gamma\ell/c$ relative to the front clock. However according to Lorentz transformations, this time shift is a relational quantity between two clocks. Einstein's approach does not tell us whether the trailing clock has shifted forward or whether the forward clock has shifted backward. On the other hand, the details of the acceleration procedure give an answer. We show that as a result of rigid acceleration, the trailing clock actually shifted backwards.\footnote{This does not mean that the clock ticks backwards in time. However, the trailing clock ticks slower during acceleration than the clock at the front, and at the end of the process it lagged behind.} But, the accelerating observer $S^\prime$ experiences a gravitational field equivalent to the acceleration and observes that the trailing clock undergoes a gravitational redshift relative to the front clock. At the end of the acceleration, the trailing clock is delayed relative to the front clock due to the gravitational redshift.
The amount of gravitational time delay of the trailing clock is larger than the time shift for the rigid acceleration and the difference is exactly equal to the special relativistic shift $\Delta t^\prime=\Delta t/\gamma=\beta\ell/c$. Therefore, the observer in the inertial frame $S_0$ observes the correct value and sign for the special relativistic time shift ; of the two clocks synchronized with respect to $S_1$, the trailing one is shifted forward by $\Delta t=\beta\gamma\ell/c$ with respect to $S_0$.\footnote{It is clear that the two clocks that are initially synchronized with respect to $S_0$ are not synchronized with respect to $S_1$ after the acceleration ends.}

 As we have discussed Einstein's approach contains some freedom regarding clock synchronization: Does the trailing clock shift forward or the clock in the front shift backward? Or do both clocks shift at certain amounts? Some authors have proposed their own synchronization hypothesis, exhausting this freedom in Einstein's approach \cite{Khan1,Khan2,Kowalski1992,Guangda,Kowalski1996,Kowalski1996-2}.
For example, according to Kowalski's synchronized clock hypothesis (SCH) spatially
separated clocks, which are synchronized in one inertial frame, maintain their synchronization in another inertial frame as long as their proper separation remains the same when they come to rest in any other inertial frame \cite{Kowalski1992,Kowalski1996}. A consequence of SCH is that there is no redshift between spatially separated clocks which are rigidly accelerating. On the other hand, if we consider the details of the acceleration procedure and derive the clock desynchronization as the result of this procedure, we realize that SCH will not be valid.

Before going into the details of our dynamical model, we will make three basic assumptions and try to elucidate their implications. These assumptions are reasonable arguments for classical rods, clocks and reference frames. Our basic assumptions are as follows: (i) {\it Simultaneous Interaction Hypothesis.} The interactions  leading to acceleration occur simultaneously at different spatial points in the accelerating frame, relative to an observer in that frame. By this we do not mean that all of the interactions occur suddenly at once and the body reaches its final velocity in one shot. But we mean that accelerating-interactions occur as a series of simultaneous interactions at different spatial points in the accelerating frame relative to an observer in that frame. We assume that the acceleration takes place by elementary particle interactions. These interactions occur simultaneously in the accelerating frame. This assumption is necessary for the acceleration to be rigid. Otherwise, the acceleration forces deform the accelerating body and change the proper lengths. For instance, assume that the accelerating body is accelerating due to electromagnetic forces. Then, from the perspective of the accelerating observer, all parts of the body interact simultaneously with the force carriers of electromagnetic interaction the {\it photons}. Of course, the simultaneity of interactions is valid up to a certain approximation precision. The interactions at the leading and trailing ends of an accelerating rod may not be exactly synchronized in the rest frame of the rod and show very small time differences. Therefore, the proper length of the rod may undergo minor oscillations. However, we assume that on average the interactions are simultaneous and the proper length of the rod does not change. (ii) {\it Acceleration Discontinuity Hypothesis.} Our second basic assumption is that acceleration is not continuous, but occurs in very small instantaneous velocity boosts. This assumption makes sense when we consider the acceleration of a classical body. According to quantum field theory interactions occur through the exchange of field quanta. Such a field exchange provides a continuous transfer of energy. Therefore, at the elementary particle level, our second assumption does not seem correct. On the other hand, the acceleration of a classical body requires numerous interactions over a relatively long time interval.
Although each interaction provides a continuous transfer of energy, each transferred energy is almost infinitesimal on the classical scale. Therefore, it would be a good approximation to assume that a classical body is accelerated by an innumerable sequence of instantaneous acceleration steps.\footnote{This is similar to the acceleration of a metal sphere by a high-intensity light beam. When a single photon interacts with a metal surface, the transferred energy is continuous; the velocity of the metal sphere changes continuously from an initial value of $v_i$ to a final value of $v_f$. But a single photon interacts in a time interval of $\Delta t=\lambda/c$, which is very tiny compared to classical scale. It is therefore a good approximation to assume that the metal sphere is accelerated by a sequence of infinitesimal acceleration steps.} (iii) {\it Constant One-way Speed Hypothesis.} We assume that the one-way speed of light is isotropic and invariant under coordinate transformations between inertial frames. The freedom of clock synchronization in relativity has been well known since Reichenbach \cite{Reichenbach} and Grünbaum \cite{Grunbaum}. By selecting different conventions for clock synchronization, the one-way speed of light can be changed without changing its two-way speed \cite{Brown2005,Anderson1998}. Therefore, by accepting assumption (iii) we actually choose the synchronization known as Poincaré-Einstein synchronization. However, it is important to note that our assumption (iii) is not completely independent of the first two assumptions. In Section 3, we will show that under assumptions (i) and (ii), there is no synchronization freedom in all other inertial frames accelerated from an initial inertial reference frame. There is freedom of clock synchronization on the initial frame of reference. However, the synchronization on the initial reference frame determines the synchronizations on all inertial frames accelerated from it.

\section{Contraction of rods and desynchronization of clocks}

Consider a rod of proper lenght $\ell$ accelerating according to our basic assumptions.
In Ref.\cite{Taylor1983} under similar assumptions, it was shown that the length of the rod observed from an inertial frame contracts by an amount exactly equal to the Lorentz contraction factor. Indeed, for simultaneous speed boosts in the accelerating frame of rod, rear boost precedes front boost in the inertial frame $S$ by a time interval $\Delta t=\beta\gamma\ell/c$. Speed of rear end of the rod exceeds speed of front end by $dv$ during $\Delta t$ and the rod is contracted by an amount $dL=dv\Delta t=\beta\gamma\ell d\beta$. Suppose the rod is initially at rest in inertial frame $S$. Then it accelerates to a final velocity $v$. The total amount of contraction is given by the following integral \cite{Taylor1983}:
\begin{eqnarray}
\label{deltalenght}\Delta L= \ell \int_0^\beta \frac{\beta^\prime}{\sqrt{1-{\beta^\prime}^2}}d\beta^\prime = \ell (1-\frac{1}{\gamma})
\end{eqnarray}
Hence, after the acceleration procedure is finished, the observer in the inertial frame $S$ 
observes a moving rod of lenght $L^\prime=\ell-\Delta L= \ell/\gamma$. Although the rear and front end points of the rod were considered during the calculation, it can be shown that a similar contraction will occur for any two points on the rod. 

For those who follow Einstein's approach, this result is one of many examples demonstrating the consistency of special relativity. According to this approach, Lorentz contraction is imposed as a rule; the laws of interactions and motion must give a result consistent with this rule. On the other hand, according to our approach, result (\ref{deltalenght}) is the {\it cause} of the Lorentz contraction: The rigid structure of the body is provided by the interactions between the particles that make up the body. If we consider the EM interaction acting on the classical scale, force carriers photons propagate at the speed of light. As a result of the speed of light being constant in different reference frames, the difference of interaction times of photons at two different points shift between these reference frames by the amount of $\Delta t=\beta\gamma\Delta\ell/c$.\footnote{Of course, we do not claim that the two ends of the rod interact. However, we assume that the contraction-causing interactions occur on a very small scale, but still on a scale for which our assumptions (i) and (ii) are valid. Accordingly, the contraction of the rod occurs cumulatively as a sum of contractions on a much smaller scale.} Consequently, the reasoning leading us to eqn.(\ref{deltalenght}) is valid and the length of the rod is contracted as given in eqn.(\ref{deltalenght}). Admittedly, there are some problems with accepting the process leading to eqn.(\ref{deltalenght}) as the cause of Lorentz length contraction. It only works for classical rods, but Lorentz contraction should also apply to elementary particles such as electrons. Can assumptions (i) and (ii) hold true at the scale of elementary particles? As we discussed earlier, according to quantum field theory, continuous energy transfer can be realized by elementary particle interactions. For this reason, the validity of assumption (ii) at the elementary particle scale requires a new hypothesis. Similarly, for the assumption (i) to be meaningful, the elementary particles must be extended objects in space. We will address these issues in the next section. Einstein's principle approach states that a moving rod contracts in length compared to its proper length. But it does not say how this happened. On the other hand, a dynamical approach provides the details.

\begin{figure}
\includegraphics[scale=1.2]{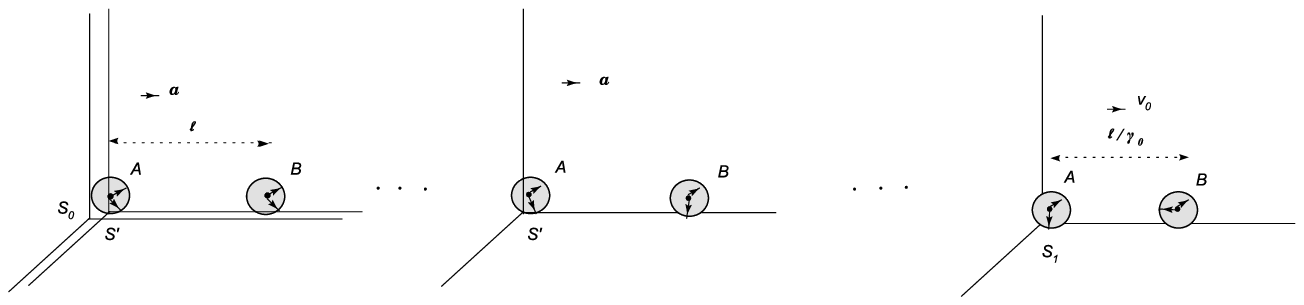}
\caption{The clocks A and B separated by a proper length $\ell$ are initially at rest and synchronized in an inertial frame $S_0$. Then, they accelerate rigidly and come to rest in an other inertial frame $S_1$ with a velocity $v_0$ relative to $S_0$. The accelerating frame of clocks is represented by $S^\prime$.\label{fig1}}
\end{figure}

Now let's come to another relativistic phenomenon for which the dynamical model provides an explanation: the desynchronization problem of moving clocks.
Consider two clocks say, A (rear clock) and B (front clock) separated by a proper length $\ell$.  Suppose clocks A and B are rigidly accelerated by small (almost infinitesimal) discrete simultaneous acceleration steps relative to an observer in the frame of the clocks $S^\prime$. The clocks are initially at rest and synchronized in an inertial frame $S_0$. Then, they accelerate and come to rest in any other inertial frame $S_1$ with a velocity $v_0$ relative to $S_0$ (Figure \ref{fig1}). According to an observer in $S_0$ frame, for each acceleration step, clock A accelerates $\Delta t=\beta\gamma\ell/c$ before clock B. Speed of clock A exceeds speed of clock B by $dv$ during $\Delta t$ and it undergoes more relativistic time dilation. Consequently, the proper time of clocks shift by
\begin{eqnarray}
\label{deltatime1}\delta \tau=\tau_B-\tau_A=(\beta\gamma\ell/c) \left[\sqrt{1-\frac{v^2}{c^2}}- \sqrt{1-\frac{(v+dv)^2}{c^2}}\right] 
\end{eqnarray}
for each acceleration step. If we expand the Taylor series and neglect the terms of order $d\beta^2$ we get
\begin{eqnarray}
 \label{deltatime2}\delta \tau=\frac{\ell\beta^2\gamma^2}{c}d\beta.
\end{eqnarray}
Integrating eqn.(\ref{deltatime2}) from 0 to $\beta_0=v_0/c$ we get the following time shift:
\begin{eqnarray}
 \label{timeshift} \Delta \tau=\frac{\ell}{c} \int_0^{\beta_0} \frac{\beta^2}{1-{\beta^2}}d\beta=\frac{\ell}{c} \left[\frac{1}{2}\ln\left(\frac{1+\beta_0}{1-\beta_0}\right)-\beta_0\right].
\end{eqnarray}
Thus, clock A has shifted {\it backwards} from clock B by $\Delta \tau$. However, this value differs from special relativistic time shift both in sign and magnitude. Indeed, according to special relativity, clock A shifts {\it forward} from clock B by proper time of $\Delta t_0/\gamma_0=\beta_0\ell/c$.

In fact, the expression (\ref{timeshift}) is compatible with special relativistic time shift. To see this, we need to consider the gravitational redshift observed by the accelerating frame of clocks $S^\prime$. Let $g$ be the constant acceleration observed in the $S^\prime$ frame of reference. The coordinate transformation between $S$ and $S^\prime$ frames is given by \cite{Wheeler}
\begin{eqnarray}
 \label{transformation} 
 &&t=\left[\frac{c}{g}+\frac{x^\prime}{c}\right]\sinh{\left(\frac{gt^\prime}{c}\right)}\nonumber \\ &&x=\left[\frac{c^2}{g}+x^\prime\right]\cosh{\left(\frac{gt^\prime}{c}\right)}-\frac{c^2}{g}\nonumber \\ &&y=y^\prime \nonumber \\ &&z=z^\prime
\end{eqnarray}
where we assume that the velocity of $S^\prime$ with respect to $S$ is along positive $x$-axis.
Therefore, the accelerating observer $S^\prime$ uses the following metric:
\begin{eqnarray}
 \label{metric} ds^2=-\left(1+\frac{gx^\prime}{c^2}\right)^2 c^2 {dt^\prime}^2+{dx^\prime}^2+{dy^\prime}^2+{dz^\prime}^2.
\end{eqnarray}
According to Einstein's equivalence principle, the observer $S^\prime$ thinks that her frame is not accelerating, but is in a gravitational field and the spacetime is described by the metric (\ref{metric}). The standard time of an observer on $S^\prime$ is coordinate dependent: $d\tau=\left(1+\frac{gx^\prime}{c^2}\right)dt^\prime$.  Without loss of generality, let's take two points $x_1^\prime$ and $x_2^\prime$ on $S^\prime$ with $x_2^\prime>x_1^\prime$ and $y_{1,2}^\prime=z_{1,2}^\prime=0$. Then we get the following gravitational time dilation formula:
\begin{eqnarray}
 \label{redshift} \frac{\Delta\tau_1}{1+\frac{gx_1^\prime}{c^2}}=\frac{\Delta\tau_2}{1+\frac{gx_2^\prime}{c^2}}.
\end{eqnarray}
Eqn.(\ref{redshift}) is an exact result for the gravitational redshift in the $S^\prime$ frame \cite{Moreau1992,Cochran1989,Landsberg1976}. Now, let's place the rear clock A at the origin of $S^\prime$, i.e. $x_A^\prime=0$. Then, the front clock B is at $x_B^\prime=\ell$. 
In this case, the following gravitational time dilation relation between clocks A and B is obtained:
\begin{eqnarray}
 \label{redshift2} \Delta\tau_B=\left(1+\frac{g\ell}{c^2}\right)\Delta\tau_A.
\end{eqnarray}
Suppose that the whole acceleration procedure takes $\tau_A$ time relative to an observer at the origin of $S^\prime$. Accordingly, at the end of the acceleration, clock B will advance by $\tau_B=\left(1+\frac{g\ell}{c^2}\right)\tau_A$. Therefore, when the acceleration procedure is complete, the observer of the rest frame of the clocks moving with the inertial frame $S_1$ will observe that clock A lags behind clock B by 
\begin{eqnarray}
 \label{redshift3} \tau_B-\tau_A=\frac{g\ell}{c^2}\tau_A.
\end{eqnarray}
Let us compare the amount of time shifts (\ref{timeshift}) and (\ref{redshift3}) between clocks A and B observed by the inertial frames $S_0$ and $S_1$. As can be seen from equation (\ref{redshift3}), clocks A and B are not synchronized with respect to $S_1$. Therefore, in order to compare the observations of frames $S_0$ and $S_1$ we must subtract (\ref{redshift3}) from (\ref{timeshift}):\footnote{According to special relativity, if clocks A and B were synchronized in frame $S_1$, the observer in frame $S_0$ would find $\Delta \tau=-\beta_0\ell/c$ instead of (\ref{timeshift}).} 
\begin{eqnarray}
 \label{timeshift2} \Delta^\prime \tau=\frac{\ell}{2c}\ln\left(\frac{1+\beta_0}{1-\beta_0}\right)-\frac{\beta_0\ell}{c}-\frac{g\ell}{c^2}\tau_A.
\end{eqnarray}
Here, $\Delta^\prime \tau$ is the proper time shift of clocks observed by $S_0$ relative to $S_1$. To simplify eqn.(\ref{timeshift2}), we make use of transformations (\ref{transformation}). From eqn.(\ref{transformation}), we get
\begin{eqnarray}
 \label{transformation2} t^\prime=\frac{c}{g} \tanh^{-1}{(\beta)} 
\end{eqnarray}
where, $t^\prime$ is the time it takes for $S^\prime$ to reach its velocity $v=\beta c$ (relative to $S_0$) with respect to a clock ticking at the origin of $S^\prime$. Since the whole acceleration procedure takes $\tau_A$ time relative to an observer at the origin of $S^\prime$
(this is also equal to the total time clock A ticked during acceleration), $t^\prime=\tau_A$ gives $\beta_0=\tanh{\left(\frac{g\tau_A}{c}\right)}\Rightarrow \ln\left(\frac{1+\beta_0}{1-\beta_0}\right)=\frac{2g\tau_A}{c}$. Hence, the first and third terms of (\ref{timeshift2}) cancel each other and we obtain the special relativistic time shift
\begin{eqnarray}
  \label{timeshift3} \Delta^\prime \tau=-\frac{\beta_0\ell}{c}.
\end{eqnarray}
The minus sign in (\ref{timeshift3}) indicates that clock A shifts forward from clock B as expected from special relativity.

Now we are in a position to discuss different clock (de-)synchronization hypotheses in special relativity. For example, according to SCH the clocks A and B which are initially synchronized in $S_0$ maintain their synchronization when they come to rest in $S_1$ \cite{Kowalski1992}. If we take the point of view of the observer $S_0$, clock A undergoes a special relativistic time shift due to acceleration; if the speed increases from $v$ to $v+\Delta v$, clock A will shift forward by $\ell \Delta v/c^2$ from clock B. Therefore, clock A advance a number of ticks $\Delta N\simeq \ell \Delta v/Tc^2$ where $T$ is the period of the clock. Consequently, during acceleration the frequency of the rear clock A increases by an amount $\Delta \nu \simeq \ell a/T c^2$ relative to that of the leading clock B as seen by the observer $S_0$ \cite{Kowalski1992}. But observer $S_0$ also observes a Doppler shift for a light wave sent from B to A. When the light wave reaches A, it is redshifted enough to compensate for the frequency shift $\Delta \nu \simeq \ell a/T c^2$ due to special relativistic clock desynchronization.\footnote{This statement is not exact, but it is valid at a certain level of approximation. In Ref.\cite{Kowalski1992} author carry all calculations only to second order in $v/c$} These two effects cancel each other. Hence, if we assume that clocks A and B are a Mössbauer receiver and source then they will resonantly interact during the acceleration. 

The described above is a brief summary of the argument of Ref.\cite{Kowalski1992}. It seems correct when we adopt the Einstein's principle approach. Indeed, we can interpret the clocks' desynchronization with acceleration as a forward shift in the time of the rear clock. On the other hand, on the basis of assumptions (i), (ii) and (iii), we show that the rear clock actually shifted backwards with respect to the front clock. But, due to Einstein’s equivalence principle the accelerating observer experiences a gravitational field and observes that the rear clock undergoes a gravitational redshift relative to the front clock. The time delay due to gravitational redshift can be thought of as a "backward time shift" of the rear clock relative to the front clock. The magnitude of this gravitational time shift  is larger than the time shift for the rigid acceleration and the difference of these two shifts is exactly equal to the special relativistic shift. Therefore, Einstein’s equivalence principle and gravitational redshift is necessary to explain special relativistic clock desynchronization. All the different (de-)synchronization hypotheses like SCH, etc., which are not compatible with the gravitational redshift observed in the accelerated frame, cannot be valid.

\section{A dynamical approach to relativity}

\subsection{The dynamical model}

In the previous section, we considered classical rods and clocks. For example, the clocks can be modeled with rotating wheels so that one period of the wheel represents a unit of time. As we have discussed in the introduction, it is a good approximation to assume
that a classical body is accelerated by an innumerable sequence of instantaneous
acceleration steps. Therefore, assumption (ii) is plausible for classical rods and clocks. Assumption (i) is valid for a rigid acceleration by definition. Moreover, rigid body kinematics is of great importance in special relativity. As Einstein said in the introductory section of his 1905 paper, special relativity theory is based on kinematics of a rigid body. This allows us to define coordinate systems on moving frames.\cite{Einstein1905,Brown2005}. If we are going to talk about observations of an accelerating frame of reference, we want the coordinate system to be maintained throughout the acceleration. In case the accelerative forces are small in relation to the cohesion forces holding the rigid rod, then it is physically reasonable to assume the acceleration is rigid and define a coordinate system on the accelerated frame. Such a coordinate system on the accelerated frame of reference can be defined  mathematically by the Fermi-Walker transport \cite{Wheeler}. On the other hand, as we will discuss in section 4, there are some limitations in this regard.

 In addition to assumptions (i) and (ii), we also accept the constancy of the one-way speed of light as a third assumption, i.e. one-way speed of light is isotropic and invariant under coordinate transformations between inertial frames. This assumption might be considered an unnecessarily strong principle that gives more than is necessary to produce the empirical content of relativity. One may also judge that our approach is a principle approach rather than a constructive one. Such judgments might be partially correct. Indeed, we do not derive constancy of speed of light from physical laws (for example from laws of electrodynamics) in a constructive 
fashion. But our approach provides a dynamic explanation for relativistic phenomena such as length contraction, time dilation and clock desynchronization. Thus, it provides answers to questions that Einstein's principle approach could not answer. For this reason, our approach also has a constructive character. It would not be wrong to say that our approach is a semi-constructive approach.

Unlike the constructive approach based on classical EM theory, which has its roots in FitzGerald, Larmor, Lorentz, Poincaré and others \cite{Bell,Brown2005,Miller2010}, our approach is inspired by quantum field theory (or more specifically, Quantum Electrodynamics, (QED)). On the other hand, if we invoke this new dynamical approach to be fundamentally valid in nature, then new hypotheses must be added to quantum field theory.
Let us postpone the discussion of this issue for a while and see the details of this new approach. In fact, in the last paragraph of the introduction, we saw the rough outline of this new approach when applied to the classical scale. According to this new approach, the interactions leading to acceleration are carried by field quanta propagating at the speed of light. We will call these field quanta {\it photons}. In the case of electromagnetic interactions, photons correspond to light quanta; usual photons.  However we will use the term photon in a more general sense. The term photon will be used to mean the field quantum propagating at the speed of light. Other properties it carries will be ignored since they are not important for our study. For example, the term photon is also used for gluons, which are the field quanta of strong interaction. 
On the other hand, strong interaction (also weak interaction) is effective on such a small scale that our assumptions (i) and (ii) do not hold (or require some additional assumptions to quantum field theory). So for now we only consider QED and photons correspond to light quanta. If we neglect gravitational interactions, QED is responsible for all interactions from the classical scale to the atomic scale, except for the atomic nucleus. 

 Let's consider a body on the scale to which our assumptions are valid. Assume that the body be a neutral or charged collection of charged particles. For example, positively charged nuclei and negatively charged electrons can come together to form atoms, and atoms can come together to form the body. This body may be part of a mechanism used to measure time. For example, it could be a periodic clock minute hand or a spinning wheel. Such periodic motion requires acceleration of the body, and acceleration occurs as a result of the interaction of the charged subparts of the body with photons. Although these interactions take place essentially at the elementary particle level, we observe that the acceleration of the body consists of small discrete acceleration steps, since a large number of charged particles interact with a large number of photons over long periods compared to the time scale of elementary particle interactions. Without loss of generality, let's think of this body as a spinning wheel.
The center of mass of the wheel is at rest with respect to an inertial frame $S$, and the observer in $S$ uses the wheel to measure time such that one period of the wheel represents a unit of time. In addition to
the spinning wheel, assume that $S$ also uses a light clock (photons traveling between two mirrors) to measure time. The mirrors are at rest with respect to $S$ and the distance between the mirrors is adjusted so that the period of the light clock is the same as that of the spinning wheel. At first, we will use the perspective of the observer in the inertial frame $S$ and discuss how all physical phenomena and events are observed by $S$. Now suppose the spinning wheel and the light clock are accelerated by the same acceleration procedure, arriving at an inertial frame $S^\prime$ with a relative velocity $v$ with respect to $S$. According to the observer $S$, the path of the photons traveling between the mirrors will be longer and the period of the light clock will increase by the usual $\gamma=\frac{1}{\sqrt{1-\frac{v^2}{c^2}}}$ factor. But can the observer $S^\prime$ determine that the period of the light clock is increased?  She can only determine this in comparison to another clock, for example she can compare the period of the light clock to the period of the spinning wheel. However, the spinning of the wheel is due to the interaction of the charged subparts of the wheel with photons. According to $S$ the trajectories of the photons deviate from their initial trajectories before acceleration, just as in a light clock. Therefore, the interaction times are delayed with respect to $S$. Accordingly, the period of the spinning wheel, whose center of mass moves with the velocity $v$, is dilated: $T^\prime=\gamma T$. Assuming that all interactions are carried out by photons, we say that the time of the $S^\prime$ frame is dilated, since this dilation occurs in the same way for all clocks of $S^\prime$. We would like to emphasize that so far we have only observed the physical world from the perspective of the observer in the inertial frame $S$. According to observer $S$, the period of the clocks in the $S^\prime$ frame increases with $\gamma$ relative to their periods before the acceleration. On the other hand, according to the first person perspective of the observer in the $S^\prime$ frame, the period of the clocks is the same as the periods observed by $S$ while the clocks were at rest with respect to S before acceleration. Therefore, according to the observer at $S^\prime$, the periods of her rest clocks are not dilated.

Now suppose a rod of length $\ell$, initially at rest in frame $S$, accelerates in the longitudinal direction and arrives frame $S^\prime$. Due to the assumption (i), the accelerating-interactions occur simultaneously in the rest frame of the rod. Since we accept the invariance of the one-way speed of light, accelerating-interactions do not occur simultaneously on the rod with respect to the observer $S$; but the rear points of the rod accelerate earlier than the front ones and move towards them with each acceleration step, getting closer. A numerical calculation based solely on the observations of the $S$ shows that the rod is indeed shortened by the $1/\gamma$  factor. An analytical calculation confirming this result is given by equation (\ref{deltalenght}). By accepting the invariance of the one-way speed of light, we actually choose a special synchronization, namely the Poincaré–Einstein synchronization. However, it is clear that any other synchronization used with assumptions (i) and (ii) cannot dynamically produce the correct contraction factor. For example, if absolute simultaneity\footnote{Some authors argue that clock synchronization is simply a matter of convention \cite{Reichenbach,Grunbaum} and a theory
maintaining absolute simultaneity is equivalent to special theory of relativity \cite{Mansouri}.} is chosen to make the time of a preferred system of reference $\Sigma$ absolute, then the one-way speed of light is not invariant \cite{Brown2005,Mansouri}. In such a case, since the accelerating-interactions will occur simultaneously in both the momentarily rest frame of the rod and the $\Sigma$ frame, the rod does not contract.\footnote{Here we implicitly assume that the length contraction is due to the time difference of the interactions taking place at different points of the rod. This does not apply when there is another dynamical cause of the length contraction. For example, as in the constructive approach based on classical EM theory, length contraction can be the result of deformation of the electric field. On the other hand, the condition of consistency of synchronization restricts us to choose Poincaré-Einstein synchronization. This issue will be discussed in detail during the comparison of our dynamical approach with the old constructive approach.} Accordingly, our dynamical approach requires a special synchronization that gives the invariance of the one-way speed of light. An important issue here is the {\it consistency of this synchronization}:
We hypothesized that all interactions are carried out by photons and that the interaction times occurring at two different locations shift according to a certain synchronization (Poincaré–Einstein synchronization) under the transformation between inertial frames. Does this shift in the interaction times cause the same amount of shift in the running clocks? The consistency of the synchronization is defined by the equality of these two shifts. In section 2, we indeed showed that the desynchronization between the clocks in the accelerating frame is exactly equal to the time shift between the photon interactions taking place at the positions of these clocks. However, gravitational time dilation must be taken into account in order to obtain the correct clock desynchronization predicted by special relativity. The conclusion we draw here is that the dynamical approach to relativity cannot be limited to special theory only, but must also include general theory.

In our dynamical approach, we assumed that accelerating-interactions occur as a series of simultaneous interactions at different spatial points in the accelerating frame relative to an observer in that frame. Such an acceleration process yields an acceleration that we define as rigid acceleration. It is also possible to consider some non-rigid acceleration procedures. One such example is the simultaneous acceleration procedure in the laboratory frame. The non-rigid acceleration procedure can be constructed artificially, for example like the one in the Dewan-Beran-Bell spaceship gedankenexperiment \cite{Bell,Dewan1959}. However, if we accept that such an acceleration procedure is originated from a fundamental dynamical law in nature, some problems arise. One problem is to find the preferred inertial frame in which the interactions will occur simultaneously. This frame becomes a preferred frame for describing the laws of dynamics in that photon interactions have a simultaneous effect on the size of body or particles with which they interact. This preferred frame also becomes the frame for which absolute simultaneity is defined. On the other hand in our dynamical approach there is no {\it particular} preferred inertial frame of reference in which the laws of physics are defined.(Or in other words, in our model there is no aether frame of reference.) But each body's or particle's rest frame is, in a sense, a preferred frame; accelerating-interactions occur simultaneously in that frame.

In the old constructive approach of FitzGerald, Larmor, Lorentz, Poincaré, Bell and others which was based on classical EM theory, the invariance of the two-way speed of light is derived constructively from the laws of electrodynamics. To be precise, the aforementioned approach starts from the assumption that the speed of light is isotropic and constant with respect to an inertial frame $S$. Under this assumption, it is shown that the volume of a moving charge observed by the observer in $S$ varies with the velocity of the charge \cite{Griffiths}. The change in the apparent volume of the electric charge is due to the retardation effects of the fields emitted from different points of the charge as it reaches the observer. Therefore, in this approach, the charges must be extended objects in space having a volume. However the size of the charges do not appear in potentials and fields. For this reason, a charge can be considered infinitesimal provided that it is not a point. The change in the apparent volume of the moving charge gives potentials known as  Liénard–Wiechert  potentials. The electric and magnetic fields can be obtained with the help of Liénard–Wiechert potentials. For example, the electric field is found to take the form \cite{Griffiths}
\begin{eqnarray}
 \label{electricfield} \vec{E}=\frac{q\vec r(1-\beta^2)}{r^3(1-\beta^2\sin^2\theta)^{3/2}}.
\end{eqnarray}
Here, $\vec r$ is the displacement vector from the charge to the point where the field is evaluated, and $\theta$ is the angle between $\vec r$ and the velocity of the charge. Thus, the electric field is distorted and loses its spherical symmetry. This form of the electric field is then applied to the charges in equilibrium and Lorentz-FitzGerald contraction is obtained. For instance, Bell applied this electric field as well as its magnetic counterpart to the dynamical equilibrium of atoms \cite{Bell}. He considered the Rutherford model of the atom and deduced the contraction of electron orbits with the help of the relativistic energy and momentum formulas that Lorentz allegedly derived empirically. Bell also showed that the orbital periods of electrons increase by the $\gamma$ factor. Thus, he concluded that the length contraction and time dilation were caused by these changes in the electron orbits of the atoms. Consider an  inertial frame $S^\prime$ moving with a constant velocity $v$ with respect to $S$. If the rulers on $S^\prime$ contract by a factor of $1/\gamma$ and the periods of the clocks increase by a factor of $\gamma$, then it is not difficult to show that the two-way speed of light does not change with respect to $S^\prime$. In fact, here we use one observer's (observer at $S$) reasoning about the observations or measurements of another observer (observer at $S^\prime$). This describes a third-person perspective: the observer at $S$ says "the observer in the $S^\prime$ frame using her contracted length and dilated time should measure that the two-way speed of light is $c$". We assume that $S$'s third-person perspective about $S^\prime$ coincides with $S^\prime$'s first-person perspective. We will discuss the validity of this assumption later. Actually, Bell also uses this assumption implicitly when transforming unprimed variables (coordinates of $S$) to primed variables (coordinates of $S^\prime$). He says:
{\it "the primed variables,...,are precisely those which would naturally be adopted by an observer moving with constant velocity who imagines herself to be at rest."}\cite{Bell}.
In the constructive approach of FitzGerald, Bell and others, the relativity of simultaneity is considered to be a convention. The relativity of simultaneity is not derived constructively; it is considered a matter of definition. This issue is one of the important differences between the old constructive approach and our approach.
In our approach, we have chosen the Poincaré-Einstein synchronization by accepting the assumption (iii). Why don't we consider synchronization as a matter of definition, but choose a special synchronization? The reason is that together with Einstein's equivalence principle, the consistency condition removes the freedom to choose the synchronization on the accelerated frame. Now, let's deduce this fact. Let's repeat our reasoning about clock desynchronization in Section 2; but assume that the synchronization shift between the positions of A and B with respect to the $S_0$ frame is not $\Delta t=\beta\gamma\ell/c$ but an unknown function of $\beta$ that we call $\alpha(\beta)$. Then, for each acceleration step the proper time of clocks shift by
\begin{eqnarray}
\label{deltatime1alpha}\delta \tau=\tau_B-\tau_A= \alpha(\beta)\left[\sqrt{1-\frac{v^2}{c^2}}- \sqrt{1-\frac{(v+dv)^2}{c^2}}\right]
\end{eqnarray}
Without loss of generality, assume that $\delta \tau$ is positive, i.e. clock A accelerates $\alpha(\beta)$ before clock B. If we expand the Taylor series and neglect the terms of order $d\beta^2$ we obtain
\begin{eqnarray}
 \label{deltatime2alpha}\delta \tau=\alpha(\beta) \beta \gamma d\beta.
\end{eqnarray}
Integrating eqn.(\ref{deltatime2alpha}) from 0 to $\beta_0=v_0/c$ we get the following time shift:
\begin{eqnarray}
 \label{timeshiftalpha} \Delta \tau= \int_0^{\beta_0} \delta \tau= \int_0^{\beta_0} \frac{\alpha(\beta)\beta}{\sqrt{1-{\beta^2}}}d\beta.
\end{eqnarray}
The quantity $\Delta \tau$ shows the difference between the proper times of A and B in the time it takes for $S^\prime$ to accelerate to a final velocity $v_0$ with respect to the observer $S_0$. Here $S^\prime$ represents the rest frame of the clocks. Note that (\ref{timeshiftalpha}) is a result obtained by an observer in the inertial frame $S_0$ about accelerated clocks. During the same acceleration process, the accelerating observer in $S^\prime$ experiences a gravitational ﬁeld and observes that the clock A undergoes a gravitational redshift relative to clock B. The amount of time shift due to gravitational time dilation is $\tau_B-\tau_A=\frac{g\ell}{c^2}\tau_A$ where $\tau_A$ ($\tau_B$) represents the total time clock A (B) ticked during acceleration. At the end of the entire acceleration process, the $S^\prime$ frame overlaps the inertial frame $S_1$ which is moving with a velocity $v_0$ relative to $S_0$. To find the total desynchronization between the frames $S_0$ and $S_1$ we must subtract the gravitational time shift from $\Delta \tau$:
\begin{eqnarray}
 \label{timeshif2talpha} \Delta^\prime \tau= \int_0^{\beta_0} \frac{\alpha(\beta)\beta}{\sqrt{1-{\beta^2}}}d\beta-  \frac{\ell}{2c}\ln\left(\frac{1+\beta_0}{1-\beta_0}\right)
\end{eqnarray}
where, we take the clock A at the origin of $S^\prime$ and use $\tau_A=\frac{c}{2g}\ln\left(\frac{1+\beta_0}{1-\beta_0}\right)$. $\Delta^\prime \tau$ shows the amount of desynchronization observed by $S_0$ between two stationary and synchronized clocks with respect to $S_1$. According to the {\it consistency of the synchronization}, the desynchronization occurring in the clocks should equal to the difference in simultaneity in the photon interactions taking place at the positions of these clocks. Therefore the consistency condition can be written as
\begin{eqnarray}
 \label{consistencycondition} \Delta^\prime \tau=-\alpha(\beta_0) \sqrt{1-{\beta_0^2}}.
\end{eqnarray}
The minus sign is due to our convention that $\delta \tau$ is positive and we multiply $\alpha(\beta_0)$ with $\sqrt{1-{\beta_0^2}}$ because $\Delta^\prime \tau$ represents the proper time shift of the clocks. If eqn.(\ref{consistencycondition}) is derived with respect to $\beta_0$ and the resulting expressions are arranged, we get the following differential equation:
\begin{eqnarray}
 \label{consistencycondition2} \frac{d\alpha(\beta_0)}{d\beta_0}=\frac{\ell}{c(1-{\beta_0^2})^{3/2}}
\end{eqnarray}
This differential equation has a solution
\begin{eqnarray}
 \label{solution} \alpha(\beta_0)=\frac{\ell\beta_0\gamma_0}{c}
\end{eqnarray}
under the initial condition $\alpha(0)=0$. As a result, it is shown that the consistency condition for synchronization gives us the Poincaré-Einstein synchronization. However, there is a subtle issue here. What we have proved is that if there is Poincaré-Einstein synchronization in the initial inertial frame $S_0$, then there will also be Poincaré-Einstein synchronization in all other inertial frames accelerated from $S_0$. But we did not prove that there must be Poincaré-Einstein synchronization on the initial inertial frame $S_0$. Indeed, if we do a resynchronization of $\tilde{t}\rightarrow t-\frac{\kappa x}{c}$ on the inertial frame $S_0$, the time dilation factor changes to $\tilde{\gamma}=\gamma(1-\frac{\kappa v}{c})$ \cite{Anderson1998}. Then instead of (\ref{consistencycondition2}) we arrive at the following differential equation:
\begin{eqnarray}
 \label{diffeqn} \frac{d\alpha(\beta_0)}{d\beta_0}=\frac{\ell}{c(1-{\beta_0^2})^{3/2}}+\frac{\alpha(\beta_0)}{(1-{\beta_0^2})}\left[\beta_0-\frac{\beta_0}{1-\kappa \beta_0}+\frac{\kappa (1-{\beta_0^2})}{(1-\kappa \beta_0)^2}\right]
\end{eqnarray}
For Poincaré-Einstein synchronization ($\kappa=0$) we obtain equation (\ref{consistencycondition2}). The solution to the equation (\ref{diffeqn}) is lengthy and we will not give it here. However, the solution depends on the $\kappa$ parameter, which indicates that the consistency condition does not determine the initial synchronization on $S_0$. By applying the consistency condition of synchronization, we have showed that if there is Poincaré-Einstein synchronization in the initial inertial frame $S_0$, then there is also Poincaré-Einstein synchronization on an inertial frame $S_1$ accelerated from $S_0$. It should be noted, however, that this proof is valid for a dynamical approach to relativity. For example, someone might argue that she can change the total gravitational shift between clocks A and B by performing a synchronization transformation such as $\tilde{t_1}\rightarrow t_1-\frac{\kappa^\prime x_1}{c}$  in the $S_1$ frame. In this case $\Delta^\prime \tau$ will be changed. We counter this argument as follows: The clocks have mechanisms that work in accordance with dynamical laws. Any desynchronization that occurs in clocks must be explainable according to the laws of dynamics. The clocks are initially at rest and synchronized in the inertial frame $S_0$. Then acceleration starts and desynchronization takes place with acceleration. The observer $S_0$ will observe that clock A lags behind clock B at each acceleration step as determined by the time dilation factor, but will not observe any other desynchronization. Similarly, according to the frame of the clocks $S^\prime$, the clocks A and B which are initially synchronized with respect to the $S^\prime$ frame, become  asynchronous according to the gravitational laws. According to the observer $S^\prime$ there is a uniform gravitational field and as a result clock A will redshift relative to clock B. An additional shift between clocks, cannot be explained by gravitational laws. In conclusion, we can say that a synchronization transformation such as $\tilde{t_1}\rightarrow t_1-\frac{\kappa^\prime x_1}{c}$ on the $S_1$ frame is unfounded as the laws of dynamics and physics.

Recall that the foundation of our dynamical approach is based on the idea that all behavior and properties of matter are the result of interactions between photons and the particles that make up matter. We think of such interactions of particles with photons as events in space-time. Thus, as a particle propagates, its world line defines a series of events connected by causal relations. These events constitute a subset of the {\it causal structure}, which is the list of all relations for all events in the universe. All observers agree on this causal structure. According to block universe picture, what is physically real in the history of the universe includes its causal structure \cite{Smolin}. What about observers and their observations? An observer, both her coordinates and her clock is a collection of the interactions of particles and, therefore, of the events they produce. The language an observer uses to express her observations, mathematics: differential equations, functions, trajectories, etc. They are emergent phenomena, in a sense they are artificial and not belong to physical reality.\footnote{We do not intend to defend the view that mathematics is the product of human imagination. But we have to proceed in this way when developing a dynamical model in line with the view that the physical world consists of causal structure.} The observer uses these concepts to organize the causal structure.
Since the causal structure constitutes physical reality, it is the same for all observers. Moreover, if two observers, for example $S$ and $S^\prime$, express their observations using the same language (organizing the causal structure using the same concepts), then $S$’s third-person perspective about $S^\prime$ must coincide with $S^\prime$’s ﬁrst-person perspective. This fact should be true for observers in all reference frames, inertial or accelerating, who use the same language and organize the causal structure using the same concepts. However, at this point, I would like to draw attention to one point at the expense of diverging little from the subject.  Although the causal structure is an immutable fact, it is possible in quantum theory for observers to have information that goes beyond this causal structure.
For example, according to an observer, the state vector may be reduced at an instant $t_A$, while for another observer it may be reduced at an instant $t_B\neq t_A$. Then, the reduction time $t_A$ ($t_B$) is information possessed by an observer that is beyond the causal structure. Another example can be given from Hardy's paradox \cite{Hardy:1992zz}. As Hardy has shown, a physical event (for example a detection event at the detector) may be carried out by a particle following a particular trajectory with respect to one observer, but by the same particle following another trajectory with respect to another observer. One observer concludes that the cause of an event is the interaction of a particle with the detector following a certain trajectory. But another observer concludes that the same event was caused by a different trajectory of the same particle. In this case, the trajectory of the particle is an information that the observer has but beyond the causal structure.\footnote{In Hardy's experimental setup, the particle's entry into the Mach-Zehnder interferometer and the particle's measurement in the detector are the two events that make up the causal structure. Between these two events, the particle is in a superposition of two separate paths (of two arms of the interferometer). No measurement is made on the superposed particle. Otherwise, the result of the experiment will change as the superposition will be destroyed. Therefore, the trajectory of the particle can be regarded as information that the observer has but beyond the causal structure. On the other hand, if we adopt a realistic interpretation of quantum theory, we can extend the causal structure by assuming that superposed particle trajectories correspond to elements of reality. Indeed, we adopt such a realistic approach. However, even if the causal structure is extended to include superposed particle trajectories, the Hardy's paradox remains unresolved. The "which-path" information that indicates which path the signal in the detector is transmitted by the particle becomes information that the observer has but beyond the causal structure. Resolution of the paradox requires the choice of "preferred frame". The subject will be discussed in section 3.2.} We can adopt an operationalist interpretation of quantum theory and argue that the information beyond the causal structure is not the element of physical reality, but merely constitutes mathematical auxiliary concepts. However, even such an operationalist approach does not provide a fully satisfactory solution to the problem. Observers are also quantum systems after all, and how do we know that they are not using this information to organize the causal structure? It cannot be guaranteed that observers $S$ and $S^\prime$ use the same language and organize the causal structure using the same concepts. Therefore, when quantum theory is included, $S$’s third-person perspective about $S^\prime$ may not coincide with $S^\prime$’s ﬁrst-person perspective. In this case, we cannot deduce the coincidence of third- and first-person perspectives, but we can accept it as an additional assumption.\footnote{In a previous article \cite{Sahin2}, we have considered a many worlds type model of quantum theory known as "Parallel Lives" and speculated a modified relativity where first- and third-person perspectives do not always coincide but can exist on separate worlds.} 

We have been particularly inspired by QED when developing our dynamical model.
However, there are strong and weak nuclear interactions that are effective in the structure of matter. As we have mentioned in the third paragraph of this section we use the term photon in a more general sense. The term photon is used to mean the ﬁeld quantum propagating at the speed of light. Other properties it carries are ignored since they are not important for our study. Gluons, the gauge bosons of the strong interaction, propagate at the speed of light. Therefore, our dynamical approach (if we neglect the smallness of the strong interaction scale) also applies to strong interactions. On the other hand, the gauge bosons $W^{\pm}$ and $Z$ for weak interaction are massive. Weak interaction is not effective in the formation of the rigid structure of matter. Thus, it does not affect our analysis in the region from classical to atomic scale. Nevertheless it is responsible for many particle decay reactions. The lifetimes of particle decays can be used to measure time. Therefore, a complete dynamical approach must also take into account weak interactions. $W^{\pm}$ bosons carry electric charge. Hence they interact with photons and receive self-energy contributions. $Z$ bosons do not have a minimal coupling to photons. But they interact 
indirectly through other particles. Each particle in the Standard Model of particle physics continually interacts with photons either directly (minimal coupling) or indirectly through other particles (radiative corrections) and receive self-energy contributions from photons \cite{Peskin,Greiner}.
For example, Higgs boson does not carry any charge. However, it interacts with photons via fermion or $W$ boson loops. Therefore, the trajectory of each particle is a series of interaction events with photons. All observers must agree on this interaction events. If we ask how the simultaneity hyperplane spanned by the interaction events for one observer would be for another observer, it is sufficient to follow the photon trajectories to find the answer. Thus, we determine the properties of matter that emerge as a result of particle interactions other than photons (such as particle decay processes), again through photon interactions. In this way we extend our dynamical model to include weak interactions.

Quantum field theory adopts a principle approach to relativity. According to this approach, it is imposed from the beginning that all interactions are Lorentz invariant. For this reason, a dynamical approach to relativity based on the conventional view of quantum field theory is not possible. On the other hand, we know that elementary particle dynamics is determined by quantum field theory. Therefore, if we want to develop a dynamical approach to relativity, we have to consider a model {\it inspired by} quantum field theory. Our model is one such candidate model that claims to provide a dynamical approach to relativity. But we should point out that if we invoke this new dynamical approach to be fundamentally valid in nature, quantum field theory must be modified. We do not see this as a flaw. On the contrary, we think that it increases the scientific value of the model because it adds falsifiability to our model. Apart from that, it should be noted that we accept a {\it realistic interpretation} of the quantum field theory. To be precise, we assume that interactions are literally carried by quantum fields. For example, we assume that the Coulomb interaction is indeed caused by photons carrying the electromagnetic force between charged particles. Such a view is in line with the perturbation theory, which is so elegantly expressed through Feynman diagrams. According to the perturbation theory, the time-ordered exponential expression of the unitary operator expands into an infinite series\cite{Peskin,Greiner}
\begin{eqnarray}
 \label{unitaryoperator} U(t,t_0)=&&T exp\left(-i\int_{t_0}^{t}dt^\prime H_{int}(t^\prime)\right)\nonumber \\
 =&& 1-i\int_{t_0}^{t}dt^\prime H_{int}(t^\prime)+\frac{(-i)^2}{2!}\int_{t_0}^{t}dt^\prime\int_{t_0}^{t}dt^{\prime\prime}T\left(H_{int}(t^\prime)H_{int}(t^{\prime\prime})\right) +...
\end{eqnarray}
where $T$ represents time-ordering and $H_{int}$ is the interaction Hamiltonian. For QED it is given by $H_{int}=\int d^3x \sqrt{4\pi\alpha}\;\overline{\psi}\gamma_\mu\psi A^\mu$. Here, $\psi$ is the fermion field, $A^\mu$ is the photon field and $\alpha$ represents the fine-structure constant. Consider a scattering process, for instance electron-proton scattering. Eqn.(\ref{unitaryoperator}) implies that the electron and proton states consist of an infinite superposition: interaction through one photon exchange (tree-level) + one photon exchange with vertex correction (one loop)+ one photon exchange with vacuum polarization (one loop) + one photon exchange with external leg corrections (one loop -unamputated) + one photon exchange with bremsstrahlung + two photon exchange + .... Experimental tests of quantum field theory are performed on a numerous number of scattered particles. For example, the luminosity of the Large Hadron Collider (LHC) at CERN is approximately $L\approx 10^{34} cm^{-2} s^{-1}$ \cite{LHC} which means that LHC can produce $10^{34}$ collisions per $cm^{2}$ and per second. For this reason, we take the average of the square of the scattering amplitude during cross section calculations. Hence, a numerous number of terms in the expansion (\ref{unitaryoperator}) contribute to the cross section. Of course, not every term contributes equally; the dominant contribution comes from the tree-level diagram and is roughly proportional to $\alpha^2$. The contributions of terms in the expansion will decrease with increasing powers of $\alpha$. On the other hand, as far as our dynamical model is concerned, we should understand what happens in a single quantum event. For a single quantum event, we should think of the expansion (\ref{unitaryoperator}) as an infinite superposition. If we can perform a measurement on these particles (such a measurement is subject to limits due to the uncertainty principle), then the superposition is reduced to one of its constituent terms with a probability proportional to some power of $\alpha$. Thus, for a single measured quantum event, only one of the terms in the superposition actually occurs. Most likely the electron and proton interact via a tree-level single photon exchange. They can also interact in another way, for example through two photon exchange or one photon exchange with one loop vertex correction etc. However, these interactions are less likely than the tree-level one photon exchange. Here we should note that we assume that the perturbation expansion (\ref{unitaryoperator}) constitutes some kind of pointer basis \cite{Zurek}; a preferred basis according to which the interactions take place relative to that basis. For example, one might think of a different expansion of the unitary operator such that it constitutes a different superposition of the state vector. But the terms in such a superposition may not contain integer number of photon fields. In such a case, we cannot claim that the interactions occur through the exchange of photon fields. Therefore, our realistic approach to quantum field theory requires the selection of a preferred basis. However, it should be noted that we do not claim that each of the Feynman diagrams in the series expansion describes physical reality exactly as described in the diagrams. What we claim is that virtual quantum fields have physical reality and that virtual fields form a preferred basis in the series expansion of the unitary operator. The series expansion of the unitary operator actually defines a superposition with each term containing an integer number of virtual quantum fields.

We have considered electron-proton scattering as an example and assumed that we could make measurements on interacted particles. What if a measurement has not been performed on the particles? In this case we are faced with numerous superpositions: Electron and proton can interact with one, two or multiple photon exchanges; their interaction strength can be modified due to vertex corrections etc. Accordingly, since particles can be in a superposition of different momenta after scattering, their orbits will be in a superposition of different world-lines in space-time. But a frame of reference, its solid axes, clocks, and any measuring device it uses is a composite structure of particles. Therefore, we must accept that the frame of reference itself, the rulers, clocks, and any measuring device it uses can also be found in a superposed state. According to this view, a frame of reference can exist in a superposition of different positions and/or velocities, its rulers in a superposition of different lengths, and its clocks in a superposition of different synchronizations and periods. But if this is true, how do classical frames of reference arise? In fact, how the classical world emerged from quantum physics has not been fully understood and is a subject that is intensively discussed today. Einselection \cite{Zurek2}, spontaneous collapse theories \cite{Ghirardi,Diosi,Penrose} are some remarkable claims to explain classicality. A promising explanation in the context of quantum reference frames is proposed in Ref.\cite{Pikovski}. It was shown in \cite{Pikovski} that time dilation causes entanglement between the internal degrees of freedom and the centre of mass of a composite particle. This entanglement leads to
decoherence of the particle’s center of mass position, if the internal degrees of freedom is traced out. It was also deduced in \cite{Pikovski} that decoherence always occurs when the center of mass of a quantum particle with internal degrees of freedom is in a superposition of two different world-lines with different total proper times. Thus, classical frames of reference may have arisen in this way.

\begin{figure}
\includegraphics[scale=1]{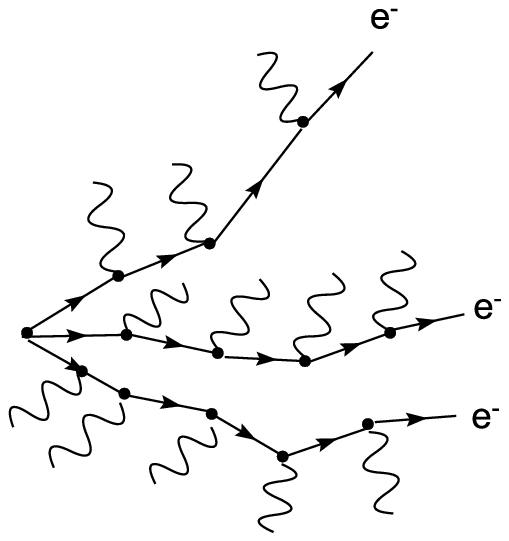}
\caption{Schematic diagram showing the superposed trajectories of the electron. Solid lines represent electrons and wavy lines represent photons. Interaction vertices are indicated by black dots. Each trajectory can be expressed as a sequence of photon interaction events in space-time.
\label{fig2}}
\end{figure}

We have mentioned that our dynamical model requires a realistic approach to quantum field theory. Such an approach may seem naive at first glance. Virtual quantum fields carrying interactions seem to originate from the approach of perturbation theory and Feynman calculus. We do not have to make the expansion of the unitary operator given by (\ref{unitaryoperator}). We could use the unitary operator without expanding it into series; Or maybe we could think of a different expansion which evolves the initial state vector into an other superposition. But if so, shouldn't we regard virtual quantum fields as auxiliary mathematical objects? In our opinion, the success of perturbation theory and Feynman diagrams is due to the fact that they contain some truths about reality that go beyond the formalism of the field theory. A similar view was expressed by 't Hooft and Veltman \cite{tHooft}. We consider virtual quantum fields to be elements of reality, even though they do not satisfy the usual dispersion relations and cannot be observed directly. On the other hand, there are some phenomena that give clues about the existence of virtual particles. Perhaps the most striking of these phenomena is the Casimir effect \cite{Casimir1,Casimir2}. The dynamical Casimir effect \cite{Casimir2} is particularly interesting. The existence of this effect was experimentally verified and the authors interpreted this effect as the conversion of virtual photons into directly observable real photons by means of a moving mirror \cite{Casimir3}. In our dynamical model, we interpret a particle's trajectory as a series of events of its interaction with photons (sometimes real but mostly virtual photons). It does not matter whether the virtual photons are (indirectly) observable or not. If the virtual photon is in exchange with an external particle that is part of a measuring device, we say that the virtual photon is observed indirectly. In this case we make a measurement on the particle. Such a measurement would be expected to eliminate the superposition  in the particle's trajectory to some extent. But if virtual photons are not observed at all, the particle trajectory remains highly superposed. An electron in the superposition of three such trajectories is given as an example in Figure \ref{fig2}. In each trajectory in superposition, the electron can interact with different numbers of virtual photons with different strengths. This is also the reason for superposed trajectories: the probability of the electron interacting with photons in different numbers and strengths causes the momentum transferred to be different. For this reason, the trajectories deviate from each other, creating a superposition. We extend the causal structure to include all photon interactions, real or virtual, observed or not. Thus, the interaction events that define the superposed trajectories are also included by the causal structure. But what if we wanted to determine the particle's trajectory and made a measurement for that purpose? In such a case, would some part of the causal structure disappear because the superposition would collapse? The answer to this question depends on the interpretation. We can assume that some part of the causal structure is deleted when the state vector is reduced and the superposition disappears. Or we can assume that some of the causal structure has ``moved out'' of our world, but within the metaverse the causal structure remains unchanged. Such an interpretation is compatible with the Parallel Lives model \cite{Brassard1,Brassard2,Waegell:2017aqh}, which provides a local and realistic interpretation of quantum theory. It could be argued that it is erroneous to extend the causal structure to include superposed particle trajectories. As we have mentioned, such an extension forces one to accept the strange partial erasure of causal structure or to adopt a many worlds type interpretation. Moreover, when the causal structure is extended, an event may have more than one cause. On the other hand, adopting a realistic interpretation of quantum theory and accepting that the state vector corresponds to a physical reality inevitably leads us to this conclusion. The important issue here is that multi causes are reduced to one after the measurement. Moreover, we have the following rationale for extending the causal structure to include superposed trajectories: It is assumed that each trajectory in superposition transforms under the Lorentz transformation just like a certain trajectory (as certain as the uncertainty principle allows) determined by measurement. The superposed trajectories must also obey Lorentz symmetry. This means that trajectories in superposition can also be studied as a sequence of events in space-time. For instance, suppose the superposed particle has some kind of "clock" internal degrees of freedom. Each tick of the clock can be thought of as an event in space-time. In this case, the particle's trajectory constitutes a sequence of ticking events. Indeed, the quantity $\int^{\tau_2}_{\tau_1}d\tau$ defines an invariant space-time distance. Here, $\tau$ is the proper time and $\tau_1$ and $\tau_2$ are two consecutive ticks of the clock. We can define the invariant $\int^{\tau_2}_{\tau_1}d\tau$ space-time distance regardless of whether the trajectory is a certain single trajectory or one of the superposed trajectories.

Our dynamical approach can be applied with good approximation from classical scale to atomic scale by simply adopting a realistic approach without changing the formalism of quantum field theory. Let's do a rough calculation to see if this is true. We consider the Coulomb interaction of an electron in an atom. 
The S-matrix element for such an interaction in QED can be written as \cite{Peskin,Greiner}:
\begin{eqnarray}
 \label{smatrix}
 S_{fi}=\frac{-ie}{\hbar}\int d^4x \;\overline{\psi}_f(x)\gamma_\mu A^\mu(x) \psi_i(x).
 \end{eqnarray}
In the lowest order of perturbation theory the initial and final states of the electron can be approximated by plane wave solutions. Since $A_0(x)$ is generated by a static point charge of $-Ze$, we have $A_0(x)=\frac{-Ze}{|\vec x|}$ and $\vec A(x)=0$. Therefore, the S-matrix element is given by
\begin{eqnarray}
 \label{smatrix2}
 S_{fi}=\frac{iZ\alpha m_e c^3}{V\sqrt{E_i E_f}}\overline{u}(p_f,s_f)\gamma^0 u(p_i,s_i)\int dt\; e^{i(E_f-E_i)t/\hbar}\int d^3x\;\frac{e^{-i(\vec p_f-\vec p_i) .\vec x /\hbar}}{|\vec x|} 
\end{eqnarray}
where $V$ represents the normalization volume and $m_e$ is the mass of the electron. $E_{i(f)}$ and $\vec p_{i(f)}$ denote the energy and momentum of the initial (final) electron. The transition probability from initial to final states is proportional to the square of the S-matrix.\footnote{The transition probability is given by $dP=|S_{fi}|^2dN_f$, where $dN_f$ is the number of final states. For a box normalization of volume  $V$, $dN_f=\frac{V}{(2\pi)^3}d^3p_f$.} In the lowest order of perturbation theory, this transition occurs via single photon exchange. The momentum transferred to the electron by the interaction is denoted by the propagator factor $\int d^3x\;\frac{e^{-i(\vec p_f-\vec p_i) .\vec x /\hbar}}{|\vec x|}$ which gives $\frac{4\pi\hbar^2}{|\vec q|^2}$; $\vec q=\vec p_f-\vec p_i$, if we take the integral over the whole space, i.e. $V\rightarrow \infty$. However, if we take the volume at the atomic scale, i.e. $V\approx a_0^3$ where $a_0$ is the Bohr radius, then we get:
\begin{eqnarray}
\label{propagator}
\int d^3x\;\frac{e^{-i \vec q .\vec x /\hbar}}{|\vec x|}=\frac{4\pi\hbar^2}{|\vec q|^2}\left[1-\cos\left(\frac{|\vec q|a_0}{\hbar}\right)\right]
\end{eqnarray}
We see from this expression that the propagator remains finite as $|\vec q|\rightarrow 0$, whereas it vanishes for $|\vec q|\rightarrow \infty$. The transition probability is proportional to the square of the propagator and the contribution to this probability for $|\vec q|\gg \frac{\hbar}{a_0}$ can be neglected.
Therefore we expect that the main contribution to transition probability comes from the region $|\vec q|\lesssim \frac{\hbar}{a_0}$. Consider an $\ell=1$ electron in an atom.
In semi-classical approximation, the momentum transferred to the electron in one revolution is about $\sim\frac{4\sqrt{2}\hbar}{a_0}$ and the time period of revolution is about $\sim 10^{-16}$s. Accordingly, it is reasonable to expect $10^{16}$ or more interactions to occur per second in an atom. Therefore, for small atom-sized pieces of  ordinary matter, we encounter photon interaction numbers as large as $10^{7}$, even in times as small as one billionth of a second. This justifies the validity of our assumption (ii). Validity of assumption (i) is due to the empirical fact that atoms and molecules and the matter they form have identical properties regardless of their previous acceleration. For example, shape-dependent properties such as dipole and quadrupole moments of atoms and molecules do not change depending on their previous acceleration. (Obviously, we assume here that the accelerating forces are very small compared to the bonding forces of atoms and molecules.) If the accelerating-interactions do not occur simultaneously on the size of the body or atoms that make up the body, then the body and atoms must flow in their rest frame and lose their shape.
In that case, the rigid structure of the body observed in it's rest frame is distorted depending on the acceleration process it has undergone. On the other hand, we would like to point out that the assumption (i) applied in a range from atomic scale to classical scale, is valid only as an average. For example, consider an accelerating rocket. As the rocket engines start to work, the accelerating-interactions will first occur at the rear end of the rocket, but will spread over the entire rocket in waves over time. Therefore, the proper length of the rocket may undergo minor oscillations. However, the interactions are on average simultaneous in the rest frame of the rocket and the proper length of the rocket does not change.

As we have shown our dynamical approach includes physically plausible assumptions for classical rods and clocks and it successfully explain special relativistic length contraction, time dilation and clock desynchronization in a dynamical way. Even for atom-sized pieces of ordinary matter and for time intervals of one billionth of a second, we can apply our dynamical approach with good approximation. The following question naturally comes to mind: Could our dynamical model be valid on a much smaller scale? For example, at the  elementary particle scale or below the elementary particle scale; perhaps near the Planck scale. If we assume that our dynamical approach is fundamentally valid in nature, then quantum field theory must be modified. First of all, for our first assumption to be meaningful, the elementary particles must be extended objects in space. This is necessary because the interactions must be able to separate the front and rear end of the particle. On the other hand, the size of the particles is not critical for our first assumption. The particles can be considered inﬁnitesimal provided that they are not a point. It is interesting that a similar feature exists in the old constructive approach. Recall that in our dynamical model, we extended space-time events to include the sequence of events that describe superposed trajectories. By doing this, we arrived at the extended causal structure. A similar extension occurs also in the momentum-energy space. Not only particle trajectories are superposed but also energy and momentum states are superposed. Since the internal energy of a composite system contributes to its mass, we must also consider superpositions of different masses. Bargmann’s superselection rule \cite{Bargmann}, which forbids the existence of superpositions of different mass states, applies only to Galilei-invariant theories. Bargmann's superselection rule is invalid in the theory of relativity. Therefore, we can consider superpositions of states with different masses. This is a known phenomenon for composite particles. However, if our dynamical model is assumed to be valid at the elementary particle level, it becomes possible for the elementary particle masses to be in a superposed state. We accept the physical reality of virtual photons in our model. Accordingly, the causal structure is extended to include self-energy interaction events. The EM field of a charged particle contributes to the particle's mass through virtual photon interactions. Since the virtual photon contributions involve a superposition of terms containing different numbers and energies of virtual photons, the state of the particle is in a superposition of states with different masses.\footnote{In the standard formalism of quantum field theory, the bare and dressed mass of an elementary particle are different from each other. However, in the standard formalism an elementary particle is assumed to have a single physical mass value, and any superposition of different physical mass states is not allowed.} We speculate that the experimentally observed masses of elementary particles arise as a result of quantum decoherence. If the number of virtual photons forming the EM field of the charged particle is large enough and the interaction time interval of virtual photons is small enough, then the decoherence time can be smaller than the time scale of the measurement. In such a case, the period of fluctuation caused by the virtual photon interactions in the internal energy remains smaller than the measurement time scale and the partial trace over the degrees of freedom of the virtual photon cloud causes decoherence. As a result, in very small time periods it is possible for elementary particles to be in a superposition of different mass states. But when a measurement is made, certain mass values (up to a certain precision) are observed.

Let's take a deeper look at our hypothesis (i) to see what other modifications to quantum field theory our dynamical model requires. Applied to elementary particles, this hypothesis requires that every point on a charged particle interacts simultaneously with the photon with respect to the particle's rest frame. If we assume that Lorentz symmetry is strictly valid in nature, then each interaction takes place simultaneously at every point on the volume occupied by the particle in space. This means that the interactions are not pointwise but occur in a volumetric region. On the other hand, if we assume that the Lorentz symmetry is not strictly valid, but is valid on average, then the interaction of the charged particle with the photon can be pointwise. In such a case, the photons interact at different points on the particle volume, but when a large number of interactions are averaged, the interactions occur simultaneously with respect to the rest frame of the particle. The hypothesis (i) is an empirical hypothesis based on the observation that the rigid structure of matter is invariant in its rest frame. Can we derive this empirical hypothesis from a more fundamental assumption about the nature of interactions? {\it We speculate that the charged particles interact with virtual photons simultaneously with respect to the center-of-mass frame of the interacting charged particles.} We will call this hypothesis $(\text{i}^\prime)$. Consider the repulsive force exerted by two electrons on each other. Since the electric field is a continuous quantity in classical EM theory, the interaction also takes place continuously. In this case, the question of whether the interactions take place simultaneously is meaningless. However, if we adopt a realistic approach to QED, we assume that interactions are literally carried by virtual photons. In this case, we can ask whether the interaction times of virtual photons with electrons are random or have a regularity to be simultaneous in a certain preferred frame. We claim that these two repelling electrons interact with virtual photons simultaneously in the center-of-mass frame of the electrons. Under the assumption $(\text{i}^\prime)$, if the interactions are averaged over long periods of time (with respect to the time periods of photon-particle interactions), composite structures composed of charged particles accelerate as a result of interactions that occur simultaneously relative to the center-of-mass frame of the composite structure. We assume that the rigid material constituting a frame of reference is such a composite structure. Thus, an observer at rest in a frame of reference is actually an observer at rest in the center-of-mass frame of the rigid composite structure forming the frame. In this way, we succeed in explaining our empirical hypothesis (i) from a more fundamental hypothesis $(\text{i}^\prime)$ about the nature of interactions. However, this explanation is still limited; It is valid when it comes to the interaction of elementary particles with each other and the composite structures they form. On the other hand, additional assumptions are needed to explain the simultaneity of the interactions in the particle's rest frame. One solution is to assume that the Lorentz symmetry is valid on average at the elementary particle level. If we accept that the elementary particles actually consist of a substructure that interact in accordance with hypothesis $(\text{i}^\prime)$, then an explanation can be provided to hypothesis (i) at a more fundamental level. Such an infrastructure should perhaps be the virtual photon cloud that contributes to the particle's self-energy. But we shall not speculate further on this matter. We do not know the formalism of the modified field theory. But we argue that a dynamical approach to relativity requires modifications to quantum field theory and sheds light on what the modifications would be. Quantum field theory adopts a principle approach to relativity. We speculate on the existence of a kind of constructive quantum field theory, that is, a quantum field theory that involves a constructive rather than a principle approach to relativity. In such a constructive quantum field theory, Lorentz symmetry should not be imposed as a fundamental principle, but should emerge from the dynamics of quantum fields.

Finally, let's discuss the validity of hypothesis (ii) on the scale of elementary particles. As we have discussed earlier, this requires a new hypothesis. But, is it a reasonable hypothesis? In quantum field theory, although free plane waves are Dirac delta normalized and their momenta can take continuous values, this feature of the theory gives an impression of approximation. For instance, instead of Dirac delta normalization sometimes box normalization is used. In the case of box normalization, the continuous integrals over momentum are replaced by discrete sums \cite{Greiner}. We then take the limit $L\rightarrow \infty$ where $L$ is the size of the box. But this requires the universe to be infinitely large, which is an approximation (although valid at a high level of precision) that is used a lot in theoretical physics. Consequently, we may expect that the momentum is discrete at a very small scale (probably near the Planck scale) below the elementary particle scale. If so, our hypothesis (ii) is valid at the elementary particle scale. Our second motivation that the acceleration should be discrete near the Planck scale is based on the algorithmic information theoretical explanation of nature. There are proposals that the universe and everything in it could be encoded as a long (perhaps infinite) string of 0s and 1s, and that the apparent laws of nature could be explained by the evolution of this long string of bits \cite{Muller,Sahin1}. If such claims are true, the smallest change in the particle's state must be described with at least 1 bit of information; states evolve discontinuously at the most fundamental level. In section 4 we will present several other arguments in support of hypothesis (ii) from the analysis of accelerated motion in the theory of relativity.

\subsection{Perspectival and dynamical effects}

{\it ``Now to the name relativity theory. I admit that it is unfortunate and has given rise to philosophical misunderstandings.`` 

Excerpt from Einstein's letter to E. Zschimmer 30 September 1921.}

It has been argued by various authors that the name "relativity" does not describe Einstein's theory correctly and that it would be more appropriate to call it the "theory of invariance" \cite{Sommerfeld,Hammond,Uehara,Sheldon}. Although Einstein's view on this issue was not very clear, he stated in a letter to Zschimmer that "invariance theory" might be a better name for his theory \cite{EinsteinLetter}. The important premise of those who think this way is that anything physically meaningful must be observer-independent. Therefore, a physical theory must be described in terms of invariant quantities. Although this point of view is compatible with the principle approach, it is not an appropriate view for a dynamical approach. In the principle approach, we think of clocks and rulers as primitive entities without structure. Each frame of reference is considered to have basic concepts of time and space (or distance), according to which events and causal structure are determined. But what about the clocks and rulers we use to determine the sequence of events? Doesn't a running clock also consist of a sequence of events? If we do not regard clocks and rulers as primitive entities without structure, then it cannot be argued that non-invariant quantities such as the periods of clocks and the lengths of rulers should not be included in a physical theory. What we mean by this is that the sentences we form using these non-invariant terms will express meaningful propositions in the context of the theory of relativity. Time and length are quantities determined by physical clocks and rulers. The change in the accelerating clock or ruler can ultimately be determined compared to other clocks and rulers in an inertial frame. Since there is an observed change, there must also be a change in physical clocks and rulers. Therefore, the change (dilation, contraction, etc.) in clocks and rulers accelerating according to an inertial frame needs to be explained dynamically. An observed change in clocks and rulers may (1) be the result of a dynamic change in clocks and rulers that we observe; (2) may be the result of a dynamic change in our own clocks and rulers.\footnote{A third possibility is that there is a dynamic change in the clocks and rulers of both our frame and the frame we observe. Such a situation may be due to the fact that both frames of reference are undergoing a real acceleration process. However, since this third case is a mixture of the other two, we will not conduct a separate analysis.}
Borrowing the terminology from Miller \cite{Miller2010}, we will refer to change or effect in type (1) as {\it dynamical}, and change or effect in type (2) as {\it perspectival}. 

We will now try to make some strong arguments to support the dynamical approach to relativity. Consider the inertial frames $S_0$ and $S^\prime$, which are initially stationary and coincident with each other. Afterwards, let an acceleration process begin for the $S^\prime$. When this acceleration process ends, $S^\prime$ moves with velocity $v_0$ relative to $S_0$. Therefore, at the end of the acceleration process, $S^\prime$ defines the inertial frame that we call $S_1$.
An observer in the $S_0$ frame observes that the rigid rods on $S_1$ contract, the clocks dilate and become desynchronized. All these effects that $S_0$ observes are dynamical. On the other hand, observer $S_1$ makes similar observations for clocks and rigid rods on $S_0$. This is obvious because Lorentz transformations are symmetrical and do not contain any history-dependent information about which inertial frame has been accelerated by a real acceleration process. However, the observations of the observer $S_1$ regarding the changes of the clocks and rigid rods on $S_0$ cannot be explained dynamically. All these effects that $S_1$ observes are perspectival. Indeed, since the observer $S_0$ does not experience any acceleration, gravitational time dilation does not occur and its clocks are not desynchronized dynamically. However, since $S_1$'s clocks are desynchronized, she compares the clocks on $S_0$ with her own clocks and concludes that $S_0$'s clocks are asynchronous. 
Similar perspectival effects occur also for length contraction and time dilation.
But one should be careful here. The fact that $S_1$'s ruler is contracted and the period of its clock is dilated does not mean that $S_1$'s (perspectival) observation of the rulers and clocks on $S_0$ should be reversed. $S_1$ does not observe that $S_0$'s ruler is expanded and the period of her clock is shortened.
Conversely, $S_1$ observes that $S_0$'s ruler is contracted and the period of its clock is dilated.  There is a symmetry between these observations of $S_0$ and $S_1$ about each other. This is a consequence of the fact that the information transfer rate for both frames is limited to the speed of light and the observed speed of light is the same in both frames. Indeed, for example, if $S_1$ wants to measure the length of a rod on $S_0$, she makes a simultaneous measurement on both ends of the rod. She can use a light signal for this measurement. Only after the light signal arrives can $S_1$ make a conclusion about the length of the rod. However, since the clocks of $S_1$ are desynchronized with respect to $S_0$, the measurement is not simultaneous for $S_0$. In addition, $S_0$ will see a delay in receiving light signals due to the movement of $S_1$. If all these details are handled carefully, it can be shown that $S_1$ will measure the length of the rod at $S_0$ to contract.

Now consider a very long rigid rod on $S_0$ with length $\ell\gg\frac{c^2}{g}$ ,where $g$ is the proper acceleration observed in the origin of $S^\prime$. We limit our analysis to an acceleration process where $g$ is constant. The rod is positioned parallel to the direction of acceleration and its front end relative to $S^\prime$ is at the origin of $S_0$. Consider a clock A placed at the origin of $S^\prime$. The total time clock A ticked during the whole acceleration process is $\tau_A=\frac{c}{g} \tanh^{-1}{(\beta_0)}=\frac{c}{2g}\ln\left(\frac{1+\beta_0}{1-\beta_0}\right)$. (See Eqn.(\ref{transformation2}) and the paragraph below.) Acceleration ends when $S^\prime$ reaches the velocity $v_0$. Immediately after the acceleration ends, the observer in the $S_0$ frame sees the $S_1$ frame moving with the velocity $v_0$ and the observer in the $S_1$ frame sees the $S_0$ frame moving with the velocity $-v_0$. Let the observer in the $S^\prime$ frame measure the length of the rod just before the acceleration begins and immediately after the acceleration is completed. By comparing these two measurement values, the observer determines that the rear end of the rod is approaching the front end with the average speed:
\begin{eqnarray}
 \label{averagevelocity}
 \bar{v}=\frac{2g\ell}{c}\frac{(1-\sqrt{1-\beta_0^2})}{\ln\left(\frac{1+\beta_0}{1-\beta_0}\right)}.
\end{eqnarray}
For $\ell\gg\frac{c^2}{g}$, $\bar v$ takes values that exceed speed of light. For instance, for $\beta_0=0.5$, $\bar v>c$ if $\ell> 4.17\times\;\frac{c^2}{g}$. The observed contraction exceeding the speed of light is puzzling but not paradoxical, since two measurements of the length of the rod were made in two different inertial frames. The initial length was measured in the inertial frame $S_0$ and the final length in the inertial frame $S_1$. On the other hand, both measurements were made by the same observer $S^\prime$. In our opinion, this fact proves that the contraction observed by $S^\prime$ is not dynamical. Indeed, the observed contraction exceeding the speed of light cannot be dynamical because it requires an infinite force.\footnote{If this very long rod were in a dynamically accelerating $S^\prime$ frame, acceleration could not occur because it would require infinite force. Dynamical constraints prohibit $g\gg\frac{c^2}{\ell}$ for proper acceleration of a rod.} Furthermore, the rod does not undergo a real acceleration process. It is the $S^\prime$ reference frame that has been subjected to a real acceleration process. Therefore, it is the clock and ruler of $S^\prime$ that changes dynamically. $S^\prime$'s observation about the contraction of the rod on $S_0$ is perspectival.

Another gedankenexperiment that constitutes an argument supporting the dynamical approach to relativity is the ''clock paradox``, or ''twins paradox``.  Consider two clocks $C_1$ and $C_2$ which are initially stationary and coincident with each other. Afterwards, let the $C_2$ clock accelerates to a velocity $v$, making a long journey at this constant velocity. Then, $C_2$ decelerates and turns back, accelerates to a velocity $-v$ and return to its original position by making a long journey at constant velocity of $-v$. Let the acceleration time intervals be negligibly small in the rest frame of the clock $C_1$. In this case, $\Delta \tau_2\approx \frac{\Delta \tau_1}{\gamma}$ can be written, where $\Delta \tau_{1(2)}$ represents the proper time of the whole trip according to the clock $C_{1(2)}$. If the fact that clock $C_2$ is accelerated by a real (dynamical) acceleration process is ignored, the observer of clock $C_2$ finds the result $\Delta \tau_1\approx \frac{\Delta \tau_2}{\gamma}$, which contradicts the observation of $C_1$. In the literature, the term ''paradox`` has been used in reference to this contradiction. On the other hand, if we take into account that the clock $C_2$ is accelerated by a real (dynamical) acceleration process and use the gravitational time dilation in the rest frame of $C_2$ as per the Einstein's equivalence principle, we get a result consistent with the observation of $C_1$ \cite{Moller}. What the clock paradox teaches us is that the clock $C_2$ is accelerated by a dynamical process, but the movement of $C_1$ relative to the frame of the clock $C_2$ is perspectival. In Feynman's own words, ''This is called a paradox only by the people who believe that the principle of relativity means that {\it all motion} is relative.``\cite{Feynman}. As Feynman pointed out, the motion of the $C_2$ clock is dynamical, but the motion of the $C_1$ clock is perspectival.

\begin{figure}
\includegraphics[scale=1]{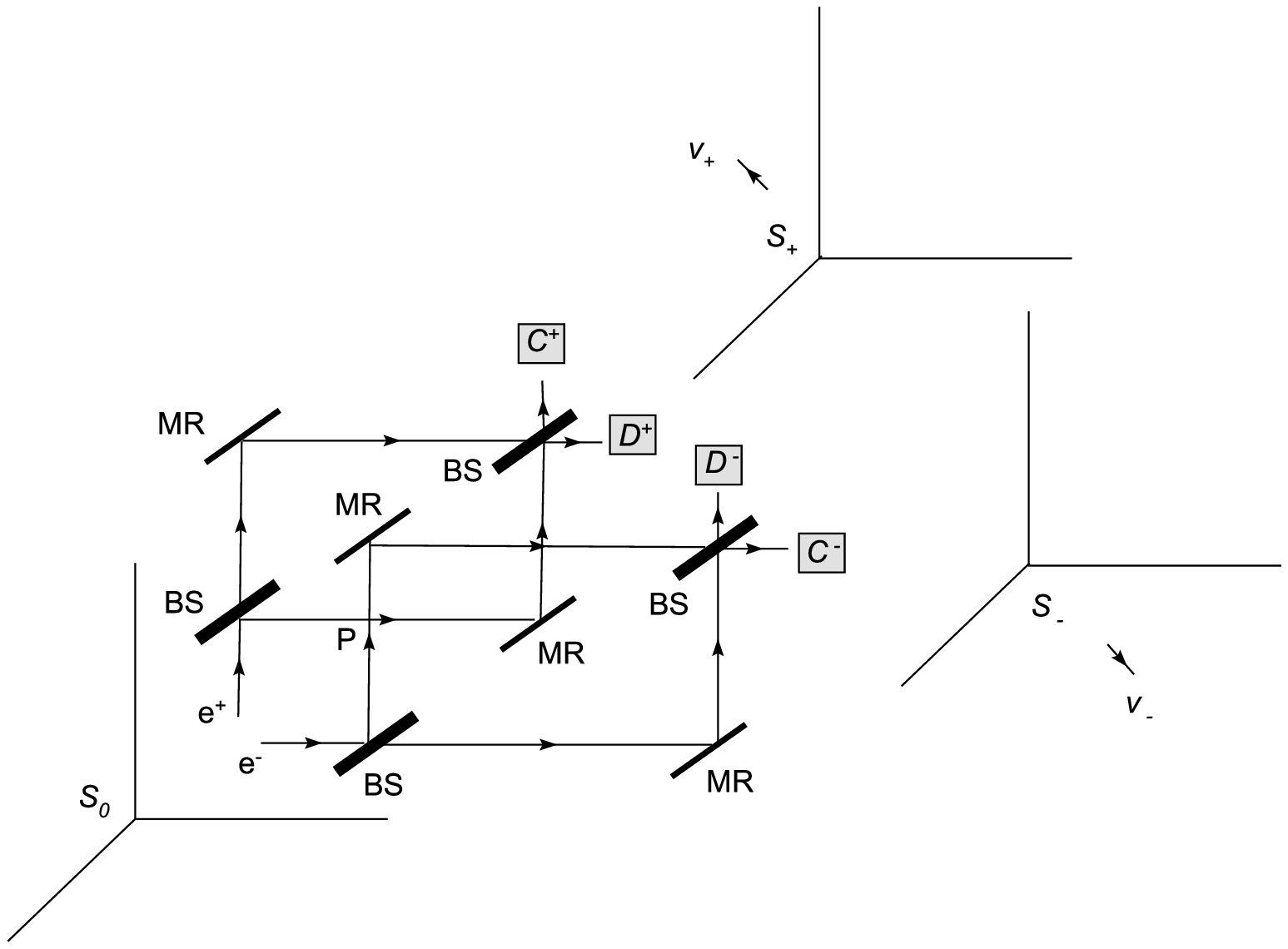}
\caption{Scheme of Hardy's experimental setup in the $S_0$ frame of reference. The BS symbol represents beam splitters and MR mirrors. $C^+, D^+, C^-, D^-$ are detectors. The inertial frames $S_+$ and $S_-$ are moving with velocities $v_+$ and $v_-$ relative to $S_0$ respectively. \label{fig3}}
\end{figure}

Finally, we will examine some of the consequences of a dynamical approach. Einstein's principle approach to special relativity deals with inertial frames moving with a constant velocity with respect to each other and the transformations between them, but ignoring the previous acceleration processes of the frames. It can be claimed that the difference of dynamical and perspectival effects is only significant during acceleration. After acceleration is completed, such a distinction is meaningless. All inertial frames are equivalent to each other regardless of previous acceleration processes. Nevertheless, although the acceleration is complete, the accelerated observer makes measurements using the clock and ruler, which have changed due to dynamical effects during the acceleration process. Therefore, her observations will continue to be perspectival. On the other hand, if we assume that the physical world is a collection of photon interaction events, and that the (one-way) speed of light is the same for all inertial frames, then the physical world produced by these interactions is the same for all inertial frames, regardless of their previous acceleration processes. So far, our dynamical model has been developed in line with this view. However, if our model is to include quantum entanglement and the consequent "spooky action at a distance" in Einstein's words, we should accept that physical reality is not solely consist of a collection of events. As an example, consider Hardy's experimental setup consisting of two intersecting Mach-Zehnder interferometers, one for positrons and one for electrons \cite{Hardy:1992zz}. The experimental setup is prepared on an inertial $S_0$ frame of reference (Figure.\ref{fig3}). Let the $S_+$ and $S_-$ frames, which are initially at rest relative to $S_0$, begin to accelerate after the setup is prepared. When the acceleration ceases, the $S_+$ and $S_-$ frames have constant velocities $v_+$ and $v_-$ relative to $S_0$. We assume that the measurements are made after the acceleration ceases. As Hardy has shown, the $v_+$ and $v_-$ velocities can be arranged so that the trajectories of the electron and positron differ in the inertial frames of $S_+$ and $S_-$ for the same measurement event in the $D_+$ and $D_-$ detectors \cite{Hardy:1992zz}. A detection event at the detector may be carried out by a particle following a particular trajectory relative to one observer, but by the same particle following another trajectory relative to another observer. One observer concludes that the cause of an event is the interaction of a particle with the detector following a certain trajectory. But another observer concludes that the same event was caused by a different trajectory of the same particle.
If the particle trajectories are assumed to be elements of reality, then one will ask what trajectories the particles actually follow. This inconsistency is known as Hardy's paradox. However we should note that the inferences of observers about particle trajectories are counterfactual. The observers do not perform any measurements to determine particle trajectories. It is known that counterfactual reasoning about the results of experiments that have not been done in quantum theory can lead us to misleading conclusions. The best-known example of this is Schrödinger's cat gedankenexperiment. It is known that it would be wrong to make a counterfactual proposition about the cat's state without measuring. Nevertheless, Hardy's paradox provides a different and much stronger argument than Schrodinger's cat gedankenexperiment. In Schrödinger's cat, it is not certain what the state of the cat will be when a measurement is made. We can only know the probabilities of the outcome. But under the conditions envisaged by Hardy, the observers' counterfactual statements about particle trajectories contain certainty. If a measurement were actually made to determine the particle's trajectory, the particle would be found with 100\% probability on the trajectory predicted by the observer's counterfactual statement.

Hardy's paradox can be applied to any realistic interpretation of quantum mechanics that relates particle trajectories to elements of reality. If we adopt a realistic interpretation of quantum theory (as in our dynamical model), then how can Hardy's paradox be resolved? The solution lies in the dynamical and perspectival distinction. The inconsistency in the Hardy's paradox about the particle trajectories actually stems from relativity of simultaneity. Hardy's setup was prepared in the inertial frame $S_0$ and remained stationary in this frame until $D_+$ and $D_-$ detector measurements were made. With respect to the $S_0$ observer, the electron and positron are measured simultaneously in the $D_-$ and $D_+$ detectors. However, according to the $S_-$ and $S_+$ inertial frames, these measurements are not simultaneous; detector measurements take place in a different time sequence.
This difference in the time order of the measurements leads to a difference in the reduction of the state vector. As a result, $S_-$ and $S_+$ frames come to different conclusions about electron and positron trajectories. On the other hand, we know that the observations of $S_-$ and $S_+$ frames about $S_0$ are perspectival. The clocks of the $S_0$ frame are not  desynchronized dynamically. Accordingly, it should be assumed that the state vector reduction takes place with respect to the $S_0$ frame. If we attribute a reality to particle trajectories, then $S_0$ must be a preferred frame. However, we would like to point out that we do not claim that $S_0$ is an aether frame in which absolute simultaneity is defined. In our opinion, the strength of special relativity and Einstein’s approach is that there is no aether frame in the universe in which the laws of physics are deﬁned. We would like to retain this feature of Einstein’s approach. On the other hand, quantum theory and the dynamical laws can give us a different preferred frame each time depending on the initial conditions. Accordingly, there is no preferential frame in the universe that can be an aether frame. But quantum theory and the dynamical laws it implies give us different frames of reference that are preferential in a limited sense (limited to some quantum phenomena). 

Let us now develop our argument by considering different variants of the gedankenexperiment. For example, suppose Hardy's experimental setup is prepared on the $S_+$ frame of reference just before $S_+$ starts to accelerate. In such a case, since $S_+$ is accelerated dynamically, the setup is exposed to accelerating-interactions. But then, the accelerating-interactions of the electron and positron with the photons have an effect similar to a measurement on the quantum system, and the preparation is disrupted. Indeed, it is reasonable to assume that the interaction of a photon  with superposed electron on two separate arms of the Mach-Zehnder interferometer will destroy this superposition. Now suppose that Hardy's setup is prepared on the $S_+$ inertial frame of reference after the acceleration is finished. There is no further acceleration process until the measurement is made. In this case, according to the time of which inertial frame does the state vector reduction occur? We know that the clocks of the $S_+$ have undergone a "real" change during acceleration, and so $S_+$'s observation about the (de)synchronization of detector measurements is a perspectival observation with respect to her altered clocks. Since the experimental setup is not subject to an acceleration, we assume that it does not matter whether it is in the rest frame of $S_0$ or $S_+$. Again $S_0$ frame is preferred; the state vector reduction takes place with respect to the $S_0$ frame. Other variants of the gedankenexperiment can also be derived. As a result, we arrive at the following conclusion:

{\it There is an initial inertial frame of reference, and this initial frame is the ''preferred`` frame for all other inertial frames that have accelerated from that initial inertial frame by a dynamical acceleration process. The term "preferred" herein means that (1) the initial synchronization of the clocks is determined in this frame; (2) the time of this initial inertial frame is considered in the state vector reduction.}

Naturally, the following questions come to mind: Is this initial inertial frame unique? If so, wouldn't that mean it's an aether frame? What determines these initial inertial frame(s)? To answer these questions, we must first clarify what we mean by a frame of reference. We do not see a frame of reference as an ancillary mathematical tool placed in space. We assume that there is an observer at the origin of the frame of reference. The observer measures time and length with a real clock and a ruler. The clock and ruler she used are not considered primitive entities without structure, but they are modeled by real running clocks and rigid material rods. When this is the case, it is asked whether the observer and the measuring devices she uses should be considered classical or quantum mechanical. We actually discussed this issue while developing our dynamical model. The frame of reference, the observer, and any measuring device (clocks, rulers, etc.) she uses is a composite structure of particles that can be found in a superposed state. Therefore, the frame of reference itself; the observer, clocks and rulers can also be in a superposed state. But then quantum decoherence happens and classicality emerges. In this paper, we have accepted that the frame of reference, observer, clock and rulers we use in the context of the theory of relativity are classical. This classical frame emerged as a result of quantum decoherence. On the other hand, if there is no concept of time independent of the clock, and the clock arose as a result of some unknown quantum decoherence mechanism, "time" must have arisen as a result of this decoherence mechanism. But there is a subtle point here. The rate of a running clock is uniquely determined by the laws of physics. For example, if we think of the clock as a spinning wheel, the period of the wheel is determined by the angular momentum and the moment of inertia of the wheel. Therefore, due to the invariance of physical laws, similar clocks for all inertial frames will have the same proper period. In other words, the laws of physics impose a superselection rule that prohibits a superposition of clocks with different time flow rates in the same inertial frame. Consequently, there is no preferred frame for the rate of flow of time, but all inertial frames are equivalent in this respect. However, the laws of physics do not impose a superselection rule that prohibits clock superposition with different phases in the same inertial frame. It seems, the choice of clock synchronization of the initial inertial frame is arbitrary in terms of physical laws.\footnote{In the context of kinematic relativity, arbitrariness in clock synchronization has been well known since Reichenbach \cite{Reichenbach} and Gr\"{u}nbaum \cite{Grunbaum}. In a dynamical theory formulated through covariant equations in the language of tensor calculus, the physical laws remain covariant under a synchrony transformation \cite{Brown2005,Anderson1998}.} The arbitrariness in clock synchronization of the initial inertial frame is exploited by quantum decoherence. Indeed, clock superposition with different phases in an inertial frame disappears by quantum decoherence, resulting in a certain clock synchronization. If we choose a convention where the one-way speed of light is $c$ in this initial inertial frame, then the one-way speed of light will be $c$ in all other inertial frames accelerated from this initial frame. However, if we believe in a classical notion of time as we have described above, then it would be pointless to talk about synchronicity before the phase of the clocks has yet emerged. According to this view, the convention of clock synchronization on the initial frame is actually the zero point calibration of clocks with different phases. To better understand this, let's consider the following gedankenexperiment: Suppose there was only light at the beginning. Then, through pair creation processes, electron-positron and quark-antiquark pairs were formed and ordinary matter came into being. The composite structure that makes up the frame of reference came together as a result of photon interactions. Afterwards, classicality emerged due to an unknown quantum decoherence mechanism. The clocks (dynamic systems that perform periodic motion) in superposition of different phases are reduced to certain initial phases. In this way, the initial frame of reference emerges. Let two such clocks emerged in different phases at two different points, A and B, on the initial frame of reference. However, in order for the A and B clocks to be used in a measurement, they must be calibrated. If we do not accept the existence of a notion of time beyond what the clock measures, then such a calibration would naturally be a synchronous convention, where the speed of light is isotropic. Of course, it is logically possible to calibrate clocks so that the speed of light is not isotropic. However, we choose to calibrate the clocks to give the simplest description of nature.

The view that time is an illusion has been claimed before and discussed in detail in the context of both relativity and quantum theory \cite{Barbour}. On the other hand, notable counter-arguments have been developed against this view \cite{Smolin}. Our approach to this issue is about questioning whether time is {\it physical} or not. If time has no physical reality, we can think of time as a property of the human mind and/or an element of the language we use to understand nature. Obviously, a physical time cannot exist in a deterministic nature. On the other hand, if there is free will, novelty takes place and physical time becomes possible. Quantum theory opens the door to free will. Quantum theory's statement about free will is perhaps most elegantly expressed in Conway and Kochen's free will theorem \cite{Conway-Kochen}. To summarize, according to the free will theorem, if the experimenter has the free will to choose the measurement to be performed, the particle also has free will in its response to this measurement.\footnote{Here we use the terminology used by Conway and Kochen in proving their free will theorem \cite{Conway-Kochen}. According to this terminology, free will is not only a property of the experimenter, but also a property of the particles or a property of the universe in a neighborhood of particles. The free will theorem is actually proved by considering spin measurements on spin-1 particles, but it can be claimed to have general validity.} The free will theorem kills determinism, but superdeterminism survives. If we interpret free will as the freedom to choose one among certain options, then we should also consider the freedom not to choose. Accordingly, the free will to choose the {\it time} of an option should also be discussed. For example, suppose the experimenter has $n$ different measurement options. Also, suppose the experimenter performs each measurement option sequentially in $N$ ($N>n$) number of identically prepared quantum systems over time.\footnote{We use the term time in different meanings in the text; sometimes in the sense of physical time, but sometimes in the sense of a feature of the human mind. Since it is not possible to reason without using the concept of time, we have to use time (with the other meaning) when describing physical time. There is no petitio principii here.} The experimenter may choose not to perform a $n_j$th measurement option at all, or perhaps choose to perform it in the $k$ th order. In this case, choosing a $n_j$th measurement option would actually be choosing it's time. What we want to say here is that free will, by definition, requires time. If there is free will in nature, then a physical notion of time also exists. As we know in quantum theory the
evolution of a closed system (involved system + environment) is unitary and deterministic. For this reason, there is no absolute physical time valid for the entire universe.\footnote{We arrive at this conclusion within the scope of our reasoning here. Of course, there are various possibilities for the existence of a universal time. For example, if the evolution of the universe is given by an incomputable 1(n)-random sequence, the universe will choose its own destiny of its own free will \cite{Sahin1}.} The free will actually emerges as a result of state vector reduction or decoherence during measurement. Therefore, it should be a feature of the subsystems of the universe. Assume that the universe as a grand quantum system evolves unitary and deterministic. When the (classical) initial frame of reference as a small subsystem of the universe emerges by quantum decoherence, a physical notion of synchronicity arises on this initial frame of reference. We say that a certain synchronicity is chosen by free will. When classicism emerges, the free will of the initial frame of reference (the free will of the observer in that frame) disappears. The observer can prepare a quantum system and can make an experiment. However, when the observer would make a measurement and what she would measure was determined as soon as the initial frame of reference emerged. In this case, we claim that {\it the particle(s) on which the measurement is made does not have the free will to choose the time of its response to the measurement.} In other words, {\it if the free will of the experimenter to choose the time of the measurement to be made disappears, the free will of the particle(s) in which the measurement is made to choose the time of the response to this measurement also disappears.} For Hardy's experiment, the freedom in the order in which the $D_+$ and $D_-$ detector measurements are made is exhausted as soon as the initial frame emerges. In this case, the physical reality of electron and positron trajectories is determined by the initial synchronization. Since physical time arises from free will, it can be argued that a statement such as "free will to choose time" contains a logical error. For those who think so, let us remind you that we use the term time in different meanings in the text (see footnote 18). Our thought is that if, as a product of the human mind, a flow abstracted from a sequence of events can be determined to some extent by free will, then such a flow must correspond to some kind of physical reality. We would also like to point out that to keep our analysis simple, we focus on the constrained case, where the initial frame of reference remains classical after its emergence. It is possible for the initial frame of reference and the observer (the observer is also considered to be the experimenter) to enter a superposed state again. For example, let's assume that the observer performs a quantum coin toss experiment using a second quantum system and performs a measurement on the first system immediately or by waiting 1 minute, depending on the result. Or imagine that the observer determines the order of the two measurements to be performed depending on the outcome of the quantum coin toss experiment. In such a case, the time of the measurement is determined by free will. The response time of the measured particle(s) to the measurement will also be determined by the same free will. In conclusion, we can say that there is no preferential absolute synchronization; but the quantum theory and the decoherence mechanism give us different synchronizations.
As long as the initial frame of reference remains its classicality, the synchronicity on the initial frame of reference is the ''preferred`` synchronicity for all other inertial frames that have accelerated from that initial inertial frame by a dynamical acceleration process.

The following question naturally comes to mind: Could two different initial inertial frames (having a relative velocity) with different synchronizations emerge from decoherence? If so, then which initial inertial frame and its synchronization should we consider "preferred"? To answer this question, we need some kind of creation model. For simplicity, let's consider the earlier gedankenexperiment: In the beginning there was only light, and all the particles that make up matter came into existence through pair creation processes... Suppose $S_0$ and $S_0^\prime$ are two different inertial frames that have emerged this way. If some of the particles that make up the $S_0$ and $S_0^\prime$ reference frames have interacted with each other in the past, then it is reasonable to assume that the interactions propagate in waves to each part in both reference frames. Here we should note that not all particles in $S_0$  need to interact with other particles in $S_0^\prime$, but they must be part of the same network of interactions. In this case, each part of the total system (combination of $S_0$ and $S_0^\prime$) is entangled with each other. Decoherence takes place in the total system and classicism emerges. According to the $(\text{i}^\prime)$ hypothesis, on average, interactions occur simultaneously with respect to the center-of-mass frame of the total system.
Therefore, the center-of-mass frame of the total system is the "preferred" frame of reference. The frames $S_0$ and $S_0^\prime$ are accelerated from the center-of-mass frame by dynamical acceleration processes.\footnote{It may be asked, according to which frame of reference do the interactions take place simultaneously? $S_0$, $S_0^\prime$ or the center-of-mass frame of the total system? For all three of these frames interactions occur simultaneously. According to the observer in the $S_0$ ($S_0^\prime$) frame, the accelerating-interactions for her frame are only a part of all interactions. These interactions occur simultaneously with respect to the center-of-mass of $S_0$ ($S_0^\prime$).} Here we implicitly assume that the total momenta of the initial photon pairs forming the total system is uniform in space. As a convention this total momentum can be chosen to be zero. The case we discussed above assumes that the frames $S_0$ and $S_0^\prime$ are in the same light cone. On the other hand, different light cones may have different "preferred" frames of reference and synchronizations. If we assume that the total momentum of the initial photon pairs is constant throughout space, then the "preferred" frames in different light cones will be at rest relative to each other. However, the "preferred" synchronization can be different in different light cones, since quantum decoherence is responsible for the initial synchronization and different light cones cannot entangle with each other. Our reasoning has been carried out within the framework of a toy cosmological model. Our aim was simply to perform a gedankenexperiment on a toy model and show how our arguments were applied. What the gedankenexperiment teaches us is that determining the family of ''preferred'' frames and synchronizations in the universe depends on the details of the cosmological model. An analysis within a realistic cosmological model requires an extensive study and exceeds the purpose of this paper.

\section{Rigid frame of reference for accelerating observers}

\subsection{Continuous acceleration versus discrete bounces}

During the derivation of the formulas (\ref{deltalenght}) and (\ref{timeshift}) for the length contraction and the time shift, we assume that each small $\Delta v$ acceleration step occurs intermittently from each other, with time intervals of at least $\Delta t=\beta\gamma\ell/c$ relative to the inertial observer $S_0$. This assumption is based on hypothesis (ii). If we abandon hypothesis (ii) and assume that acceleration occurs continuously, then under what conditions the formulas (\ref{deltalenght}) and (\ref{timeshift}) are obtained? In this subsection, we will seek an answer to this question and provide some arguments in support of the validity of hypothesis (ii).

Consider the acceleration of clocks A and B depicted in Figure \ref{fig1}. But now let's assume that the acceleration is continuous and the instantaneous proper acceleration is constant $g$. We will show that the following two cases yield different results: (1) The case in which the acceleration process is completed. (2) The case in which the acceleration process is ongoing. In case (1), the acceleration process ends and the clocks come to rest in the inertial frame $S_1$ moving with a velocity $v_0$ relative to $S_0$. In case (2), we are at a moment when the acceleration process continues. In this case, we consider the momentarily rest frame of the clocks.

{\bf Case (1): Completed acceleration}

Since the proper acceleration is constant, the hyperbolic transformations (\ref{transformation}) can be applied between the frames $S_0$ and $S^\prime$.
The time it takes for the $S^\prime$ frame to reach its final $v_0$ velocity is $t_0^\prime=\frac{c}{g} \tanh^{-1}{(\beta_0)}$. By substituting this expression in equations (\ref{transformation}), the times and positions of clocks A and B when they reach the final velocity $v_0$ according to the frame of reference $S_0$ are found as follows:
\begin{eqnarray}
 \label{t0times}
 \nonumber
 &&t_{0A}=\frac{c}{g}\sinh(\tanh^{-1}{(\beta_0)})=\frac{c}{g}\beta_0 \gamma_0\\
 &&t_{0B}=\left(\frac{c}{g}+\frac{\ell}{c}\right)\sinh(\tanh^{-1}{(\beta_0)})=\left(\frac{c}{g}+\frac{\ell}{c}\right)\beta_0 \gamma_0
\end{eqnarray}
\begin{eqnarray}
 \label{x0positions}
 \nonumber
 &&x_{0A}=\frac{c^2}{g}\cosh(\tanh^{-1}{(\beta_0)})-\frac{c^2}{g}=\frac{c^2}{g}(\gamma_0-1)\\
 &&x_{0B}=\left(\frac{c^2}{g}+\ell \right)\cosh(\tanh^{-1}{(\beta_0)})-\frac{c^2}{g}=\left(\frac{c^2}{g}+\ell\right)\gamma_0-\frac{c^2}{g}
\end{eqnarray}
We see from these expresions that
\begin{eqnarray}
 \label{deltat0}
 \Delta t_0=t_{0B}-t_{0A}=\frac{\ell}{c}\beta_0 \gamma_0
\end{eqnarray}
and
\begin{eqnarray}
 \label{deltax0}
 \Delta x_0=x_{0B}-x_{0A}=\ell \gamma_0 .
\end{eqnarray}
We see from eqns.(\ref{t0times}) and (\ref{deltat0}) that the clocks A and B reach the velocity $v_0$ at different times with respect to $S_0$. The reason for this is that the accelerations of the clocks A and B are different relative to $S_0$. This is a result of rigid acceleration. Therefore, eqn.(\ref{deltax0}) does not give the correct Lorentz contraction factor as it shows the distance between positions observed at different times. Let's consider the clocks A and B as the left and right ends of a rigid rod of proper length $\ell$. A reaches its final velocity $v_0$ at time $t_{0A}$ and its acceleration ends. From now on, it moves with a constant velocity $v_0$. B reaches the velocity $v_0$ at a later time $t_{0B}$. Since A will move at a constant velocity during the time period $\Delta t_0$, at a moment after acceleration is complete ($t\geq t_{0B}$), the distance A-B is given as:
\begin{eqnarray}
 \label{deltaxAB}
 \Delta x_{AB}=\Delta x_0 - v_0 \Delta t_0 =\frac{\ell}{\gamma_0}.
\end{eqnarray}
This is the A-B distance on the inertial frame $S_1$. The correct Lorentz contraction value is obtained. 

Now let's calculate the time shift observed between clocks A and B of frame $S_1$ relative to an observer in the $S_0$ frame. According to an observer at $S_0$, the total proper time clocks A and B ticked during their acceleration from 0 initial velocity to $v_0$ final velocity are given by
\begin{eqnarray}
 \label{tau0A}
 &&\tau_{0A}=\int_0^{t_{0A}} \sqrt{1-\frac{v_A^2}{c^2}} dt_A= \frac{c}{2g}\ln \left[\frac{1+\beta_0}{1-\beta_0}\right]\\
 \label{tau0B}
 &&\tau_{0B}=\int_0^{t_{0B}} \sqrt{1-\frac{v_B^2}{c^2}} dt_B=\frac{1}{2} \left(\frac{c}{g}+\frac{\ell}{c}\right)\ln \left[\frac{1+\beta_0}{1-\beta_0}\right].
\end{eqnarray}
We see from eqns.(\ref{tau0A}) and (\ref{tau0B}) that the clocks shift by $\tau_{0B}-\tau_{0A}=\frac{\ell}{2c}\ln \left[\frac{1+\beta_0}{1-\beta_0}\right]$. On the other hand, since clocks A and B reach their final velocity $v_0$ at different times, this time shift is not the one observed by observer $S_0$ between the clocks of frame $S_1$. Since clock A completes its acceleration before clock B, it ticks by $\Delta t_0/\gamma_0$ during $\Delta t_0$. Therefore, the observer at $S_0$ will observe the following time shift between clocks A and B of the inertial frame $S_1$:
\begin{eqnarray}
 \label{deltatauAB}
 \Delta \tau_{AB}=\tau_{0B}-(\tau_{0A}+\frac{\Delta t_0}{\gamma_0})=\frac{\ell}{2c}\ln \left[\frac{1+\beta_0}{1-\beta_0}\right]-\frac{\ell \beta_0}{c}.
\end{eqnarray}
This time shift coincides with (\ref{timeshift}).

{\bf Case (2): Ongoing acceleration}

At a moment when acceleration continues, the velocities and accelerations of points A and B are different with respect to $S_0$. Indeed, it is clear from (\ref{transformation}) that
\begin{eqnarray}
 \label{xversust}
 x=\sqrt{\left(\frac{c^2}{g}+x^\prime\right)^2+c^2t^2}-\frac{c^2}{g}
\end{eqnarray}
and
\begin{eqnarray}
\label{vversust}
v_A(t)=\frac{dx}{dt}|_{x^\prime=0}\;,\;v_B(t)=\frac{dx}{dt}|_{x^\prime=\ell}.
\end{eqnarray}
Thus, according to an observer in the $S_0$ frame, the clocks A and B are never at rest relative to each other in any in instant during acceleration. For this reason, the $S^\prime$ frame is not well-defined for an observer in the $S_0$ frame of reference. Therefore, the predictions for length contraction and clock desynchronization become ambiguous. Is frame $S^\prime$ the stationary frame of clock A? Or is it the stationary frame of the clock B? This might be considered as an spurious discussion. But it is not so, because it is an important discussion in determining both length contraction and clock desynchronization. Indeed, what velocity value should be used in the calculation of the $\gamma$ factor? $v_A$ or $v_B$? Similar ambiguity exists in the calculation of clock desynchronization. If we choose the $S^\prime$ frame as the momentarily rest frame of one of the clocks, it can be shown that results obtained for length contraction and clock desynchronization are inconsistent with the predictions of special relativity. On the other hand, if $\ell$ is infinitesimal then $v_A=v_B$ can be taken and the frame $S^\prime$ can be defined. In such a case, it can be shown that the formulas (\ref{deltalenght}) and (\ref{timeshift}) are verified (Appendix A).

\begin{figure}
\includegraphics[scale=1]{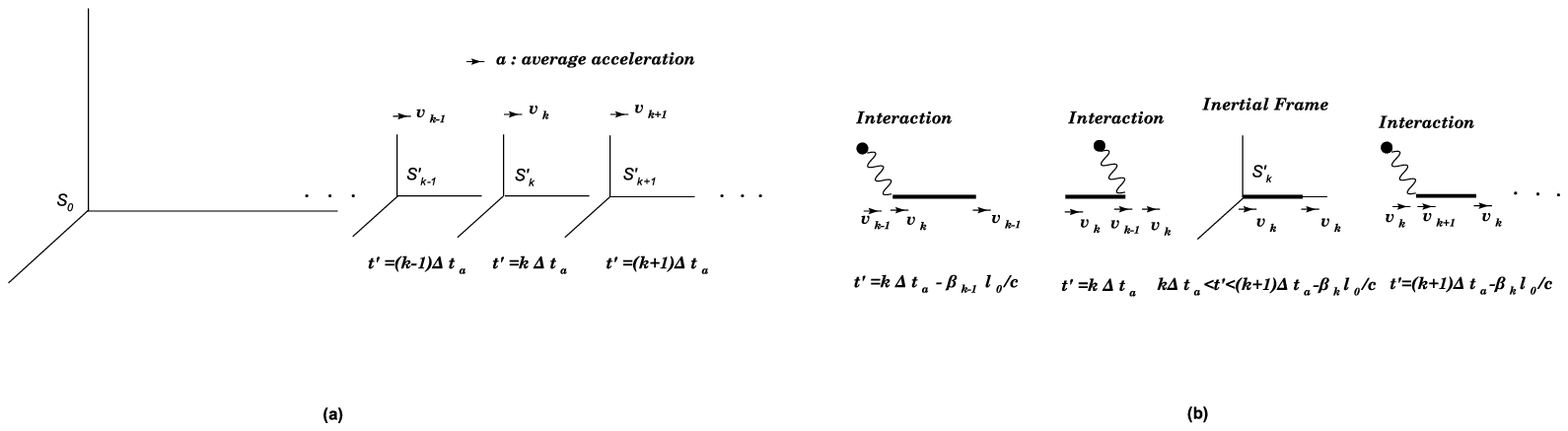}
\caption{Panel (a) shows inertial $S^\prime_k\;;k=1,2,..,N$ frames immediately after sequential jump steps. $t^\prime$ represents the proper time of the accelerating $S^\prime$ frame. Panel (b) shows the rear and front ends of the rigid rod bouncing at $\Delta v$ speed by interaction. The figures are presented according to the observation of the observer $S_0$.\label{fig4}}
\end{figure}

According to the $S_0$ frame, it is not possible to define the momentarily rest frame of the rigidly accelerated clocks A and B with finite proper length, but such a frame can be defined according to an observer on one of the clocks. Indeed, imagine that there are rockets adjacent to clocks A and B, and that the clocks are accelerated by the operation of rocket engines. In addition, let's assume that the amount of fuel that rocket engines burn per unit time and thus the acceleration they provide can be adjusted by a mechanism. Rocket mechanisms can be adjusted so that an observer at A observes that B's distance does not change during acceleration. Such an acceleration would be a rigid acceleration, and a frame of reference $S^\prime$ can be envisaged in which the clocks A and B are momentarily at rest at each moment of the acceleration. However, for a given $g$, the length $\ell$ cannot be very large, but must obey the constraint $\ell<\frac{c^2}{g}$ \cite{Wheeler}. Otherwise, the acceleration of A will be infinite. Let $\ell$ be a sufficiently small finite length that fits this constraint. In this case, a family of inertial frames $S^\prime$ can be defined in which $\ell$ is momentarily at rest with respect to any observer located at a point on $\ell$ (for example, with respect to A at the left end of $\ell$ or B at its right end). On the other hand, according to the observer at $S_0$, it is not possible to define such a family of inertial frames. So let's ask the question: Can a sufficiently small but finite rigid frame of reference be defined for an accelerating observer? We know that such an infinitesimal frame can be defined. But "infinitesimal" is an idealized definition. If we accept that there is no infinitesimal length scale in nature, then it is reasonable to assume that an accelerating reference frame with a very small but finite scale 
$\ell_0$ can be defined. Naturally, the $\ell_0$ scale can be associated with the Planck scale, i.e. $\ell_0\simeq \ell_P$. If it is possible to define a rigid frame of reference covering a finite domain at the scale $\ell_0$ for the accelerating observer, then how can this result be reconciled with $S_0$'s observation? The way to achieve this is to take the acceleration in discrete bounces, with each step applied to the $\ell_0$ scale a {\it completed acceleration}. In this case, relative to the accelerating observer, acceleration occurs in simultaneous jumps at both ends of $\ell_0$ (and at every point of $\ell_0$, if there is a scale smaller than $\ell_0$). Accordingly, $\ell_0$ is at rest in some inertial frame immediately after each jump, and the proper length $\ell_0$ does not change during the entire acceleration process. On the other hand, according to the observer in the inertial frame $S_0$, the rear end of $\ell_0$ jumps early from its front end by the $\Delta t=\beta\gamma\ell_0/c$ time slot at each step. If we assume that the acceleration takes place by instantaneous $\Delta v$ steps with the $\Delta t_a$ time intervals in the accelerating frame of reference, then we impose the condition $\Delta t_a\geq \Delta t/\gamma =\beta\ell_0/c$. For $0\leq \beta<1$ we get $\Delta t_a\geq \ell_0/c$. Under this condition, immediately after each acceleration step, the rigid axes defining the accelerating coordinate system will be at rest with respect to some inertial frame (see Figure \ref{fig4}).

\subsection{The equivalence principle}

\begin{figure}
\includegraphics[scale=1]{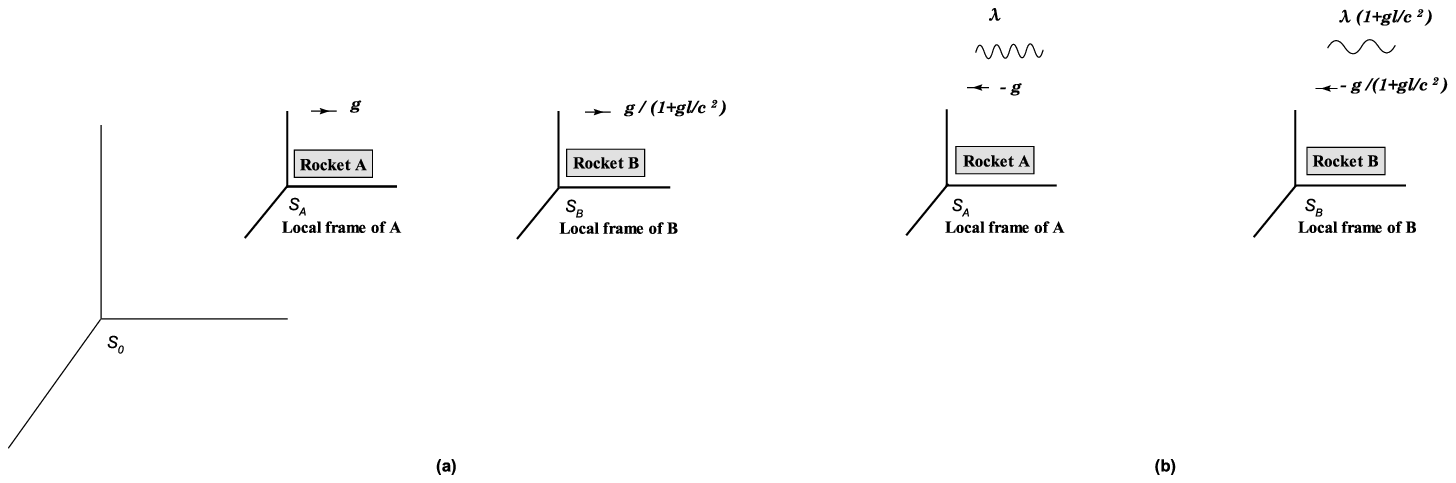}
\caption{Panel (a) shows the acceleration of rockets relative to the inertial observer $S_0$. $S_a$ and $S_b$ represent local frames attached to rockets A and B. $S_a$ and $S_b$ experience proper accelerations $g$ and $\frac{g}{1+\frac{g\ell}{c^2}}$, respectively. Panel (b) shows the same acceleration process relative to $S_a$ and $S_b$ frames. An observer at $S_b$ observes that a light wave of wavelength $\lambda$ emitted from A is redshifted by $\lambda (1+\frac{g\ell}{c^2})$.\label{fig5}}
\end{figure}

Let's try to understand the implications of the existence of a very small but finite, accelerating rigid frame of reference regarding the equivalence principle.
Consider the gedankenexperiment in the last paragraph of section 4.1. The rocket engines at points A and B are fired in such a way that the proper A-B distance $\ell$ remains unchanged. This gives us a rigid acceleration,
and with respect to observer A or B the accelarating frame of reference $S^\prime$ can be envisaged in which the observers A and B are momentarily at rest at each moment of the acceleration. Suppose observers A and B have accelerometers that can measure the proper acceleration caused by the rocket engines. It can be easily shown from the equations (\ref{xversust}) and (\ref{vversust}), that the momentum of the rockets are
\begin{eqnarray}
 p_A=mgt\;,\;\;p_B=\frac{mgt}{1+\frac{g\ell}{c^2}}
\end{eqnarray}
with respect to the inertial observer $S_0$. Here we neglected the decrease in the mass of rockets during acceleration. Since the momentum transferred per unit time is equal in the frame $S_0$ and the instantaneous inertial frames of the rockets, i.e. $\frac{dp_{A(B)}}{dt}=\frac{dp^\prime_{A(B)}}{dt^\prime}$, the proper accelerations observed by the observers in rockets A and B are:
\begin{eqnarray}
\label{properaccelerations}
 g_A=g\;,\;\;g_B=\frac{g}{1+\frac{g\ell}{c^2}}.
\end{eqnarray}
As can be seen from above, proper accelerations are different for observers in local reference frames on rockets A and B (Figure \ref{fig5}-(a)). In the beginning, we designed the gedankenexperiment for an acceleration process in which rocket engines are fired in such a way that the A-B proper distance remains unchanged. But if observers A and B are measuring different accelerations, don't they conclude that the proper distance A-B will change over time? Doesn't this fact contradict the initial assumption of the gedankenexperiment? No, there is no point in conflict with the initial assumption. The answer lies in gravitational time dilation. The observer in rocket B will observe a redshift in the energy of A by a factor of $\frac{1}{\left(1+\frac{g\ell}{c^2}\right)}$ (Figure \ref{fig5}-(b)). Conversely, the observer in rocket A will observe a blue shift in B's energy by a factor of $\left(1+\frac{g\ell}{c^2}\right)$. Indeed, when a photon emitted from A reaches B, observer B measures a redshift in the wavelength of the photon. Suppose rockets are accelerated via photon emission and by the impulse generated by photon momentum. Then, observer at $S_B$ observes that the energy gained by rocket A decreases and consequently its acceleration decreases to $\frac{g}{1+\frac{g\ell}{c^2}}$. This is equal to the proper acceleration of B. Therefore, according to observer $S_B$ the accelerations of rockets A and B are equal to $\frac{g}{1+\frac{g\ell}{c^2}}$. On the other hand, observer at $S_A$ observes that the acceleration of rocket B increases due to blue shift. According to observer $S_A$ the accelerations of rockets A and B are equal to $g$. As a result, according to the observer on both rockets, the accelerations of the rockets are equal and the proper distance $\ell$ remains constant during the acceleration.

We have seen that such a rigid reference frame is always definable with respect to an accelerating observer A or B, provided that $\ell<\frac{c^2}{g}$. Similarly, we can think of each point between rockets A and B as material points modeled by accelerating rockets. Thus we can construct a rigid rod connecting points A and B. Rigid rods constructed in this way can be used to model the axes of the accelerating $S^\prime$ frame. However, defining an accelerating rigid frame of reference covering such a finite region with respect to the inertial observer $S_0$ is problematic. As we have already shown, such an accelerated frame does not maintain its rigid structure with respect to the inertial frame $S_0$, but it flows. In the context of classical relativity, unless we abandon the assumption of the continuity of acceleration, such an accelerating rigid frame can only be defined with respect to an inertial frame if its size is infinitesimal. Some authors have ignored this fact and claimed that the equivalence principle is invalid \cite{Desloge}. The point that is missed in such assertions, is that a rigid acceleration is not uniform with respect to an inertial frame. Indeed, relative to the inertial frame $S_0$, the acceleration of the frame $S^\prime$ is not uniform, but different points on its axis have different accelerations. Therefore, the equivalence principle can only be applied in the infinitesimal neighborhood of a point of the $S^\prime$ frame. In such an infinitesimal neighborhood, the equivalence principle is unambiguously valid. 

For the Kottler-M\o{}ller metric (\ref{metric}) describing the rigidly accelerating $S^\prime$ frame, the Riemann curvature tensor is $R_{\beta\sigma\mu\nu}=0$. Therefore, it can be concluded that the metric (\ref{metric}) does not describe a gravitational field. On the other hand, although the metric (\ref{metric}) does not represent a uniformly accelerating frame, it converges to a metric describing uniform proper acceleration in an infinitesimal neighborhood. The metric representing such a uniform proper acceleration is given by \cite{Desloge}
\begin{eqnarray}
\label{metric2}
 ds^2=-e^{(2g x^\prime/c^2)} c^2 {dt^\prime}^2+{dx^\prime}^2
\end{eqnarray}
where the coordinates $y^\prime$ and $z^\prime$ perpendicular to the acceleration direction are neglected. This metric reduces to the Kottler-M\o{}ller metric (\ref{metric}) in the limit $x^\prime\rightarrow0$. It can be shown that the Riemann curvature tensor for metric (\ref{metric2}) is nonzero:
\begin{eqnarray}
 R_{1001}=R_{0110}=\frac{g^2}{c^4}e^{2g x^\prime/c^2}=-R_{1010}=-R_{0101}
\end{eqnarray}
Therefore it describes a gravitational field. What we can say about the (\ref{metric}) metric is that it describes the gravitational field locally, but not globally.  

We have seen that gravitational redshift occurs in a way that compensates for the difference between the observations of accelerated and inertial observers. The inertial frame $S_0$ observes that the accelerating $S^\prime$ frame is flowing. According to $S_0$, more energy is transferred to the fast-flowing regions of $S^\prime$ than to the slow-flowing regions. But the $S^\prime$ frame does not observe such a flow due to the nature of rigid acceleration. Thus, the energy supplied by the flow is not observed in the accelerating frame, which leads to a redshift in energy. Therefore, the "flow" observed by the inertial frame is the origin of the gravitational redshift. 
In the context of classical relativity, this flow is considered to be a continuous quantity. However if we take the acceleration as discrete bounces, as we argued in the previous subsection, the flow becomes discontinuous. Suppose the acceleration occurs in instantaneous $\Delta v$ steps with time intervals $\Delta t_a$ satisfying the condition $\Delta t_a\geq \ell_0/c$ in the accelerating reference frame (Figure \ref{fig4}). In such a case, since the acceleration on the time scale $\Delta t_a$ is a completed acceleration, regions of size $\sim\ell_0$ on the flow are represented by inertial frames. Accordingly, if we restrict ourselves to this time scale, gravitational redshift and other general relativistic effects do not arise, but we remain in a world where only special relativity applies.  
Indeed, for an observation of the $S_0$ inertial frame in the time interval $\Delta t_a$, there is no change in the velocity of the flow, so there is no energy transfer to different regions of the flow. As a result, no gravitational redshift occurs.\footnote{Another way to see this is to consider the Doppler shift observed by $S_0$ between two separate points on the accelerating frame. Consider two points on the accelerating frame, spaced $\ell \lesssim \ell_0$. In this case, when a light wave emitted from the rear point reaches the front point, the wavelength does not undergo a shift due to acceleration. This is because the light wave reaches the front point before the $\Delta v$ bounce occurs.} On the other hand, for a long time interval such that $\Delta t >>\Delta t_a$, a large number of consecutive acceleration steps ($\Delta v$ bounces) occurred. In this case, the acceleration becomes continuous as it is averaged over many small discrete acceleration steps. We argue that gravitational redshift and general relativity emerge in this way, that is, the spacetime continuum exists only in the average sense. We admit that our claim that general relativity emerged in this way is a speculation in need of further arguments. However, when we consider that the equivalence principle is a corner stone of general relativity, we dare to make such a speculation.
Obviously, the equivalence principle is a necessary but not a sufficient principle for general relativity \cite{CliffordWill}. The equivalence principle provides a local equivalence of the acceleration and the gravitational field. On the other hand, the equivalence principle is insufficient to determine the global properties of the spacetime manifold.
Nevertheless, our dynamical model predicts some implications for general relativity. If acceleration is of a discontinuous nature, the application of the equivalence principle implies that the gravitational field is also discontinuous.
For local flatness of the spacetime manifold, an infinitesimal region is not needed, but in a very small but still finite region of size $\sim\ell_0$ the metric reduces to the Minkowski metric. In regions of size $\ell>>\ell_0$, we speculate that the spacetime metric evolves from the Minkowski metric by gradually changing in small discrete steps. 


\section{Conclusions}

In this paper, we have approached the theory of relativity within the framework of a dynamical model that we developed inspired by quantum field theory. Our approach provides a dynamical explanation for relativistic phenomena such as length contraction, time dilation and clock desynchronization, which Einstein's principle approach leaves in the dark. It should be noted that our approach differs from that of Lorentz and others \cite{Prokhovnik}, which includes aether models.
In our view, the strength of Einstein's approach is that there is no single preferred frame of reference in the universe that can be called the aether frame in which the laws of physics are defined. On the other hand, when quantum theory is involved, we should be open to the fact that different frames of reference may temporarily have the status of preferred frame. We have argued that there is an initial inertial frame of reference, and this initial frame is the ”preferred“ frame for all other inertial frames that have accelerated from that initial inertial frame by a dynamical acceleration process. On the other hand, this initial frame of reference is not unique, but different initial frames of reference may exist at different points in spacetime. If we consider the initial reference frame as a classical frame, the initial synchronization on it is endowed by quantum decoherence. In other words, the universe freely chose the "preferential" synchronization on this initial frame of reference. In the paper, we also use the term "preferred" in a second meaning. Our simultaneous interaction hypothesis (hypothesis (i)) implies that the center-of-mass frame of the accelerating system is in some sense a preferred frame. This is in the sense that the interactions occur simultaneously in this frame and is in fact necessary for rigid acceleration to take place. We then claimed that hypothesis (i) can be derived from the more fundamental hypothesis $(\text{i}^\prime)$, which states that interacting particles interact simultaneously with force carriers in the center-of-mass frame of the interacting particles. Quantum field theory deals with relativity from the perspective of the principle approach. Therefore, the question remains unanswered as to which frame of reference the fundamental interactions occur simultaneously. For someone who takes the principle approach strictly, this might even be a meaningless question. However, within the framework of our dynamical approach, this question is a meaningful question and requires that the center-of-mass system of the interacting particles is a preferred frame. It can be said that nature chooses this frame as the frame in which the interactions take place simultaneously and makes it preferred. It should be noted that the notions of preferred frames in our model do not violate the principle of relativity.\footnote{As we have discussed before, the choice of clock synchronization of the initial inertial frame is arbitrary in terms of physical laws. On the other hand, the initial frame is also the preferred frame for state vector reduction. But the preferentiality here is related to a part of reality that we do not have access to and does not lead to a change in the measurement results. Accordingly, the principle of relativity is not violated. Nevertheless, if there are some laws of physics that determine the reduction of the state vector, it can be said that the principle of relativity is violated for these laws. The notion of preferentiality, which we use for the frame of reference in which interactions are simultaneous with respect to it, is also not contrary to the principle of relativity. Such a preferentiality property is Lorentz invariant, although not manifestly Lorentz invariant.} But it requires an extension of the language of relativity theory.\footnote{If we consider the theory of relativity as an axiomatic system, we can imagine that we form sentences with grammatical rules from the alphabet of this system. A strict adherent of the principle approach might argue that sentences that make sense in the context of relativity theory should contain only invariant quantities, and that, if we make a formalization of relativity, grammatical rules should take this fact into account. According to this view, sentences such as "this rod is 3 cm long" or "these two events happened simultaneously" are meaningless in the context of relativity since they are observer dependent. Thus, it can be said that if a complete formalization is made for relativity, grammatical rules should be arranged in such a way that they do not allow such sentences. We disagree with this view and argue that the language of relativity theory should be extended to include such sentences.} We make the conjecture that quantum field theory based on a principle approach to relativity should be replaced by a constructive quantum field theory to ground the dynamical model we propose. 
We do not know the formalism of this constructive quantum field theory. However, we have made predictions about several features of this theory based on our dynamical model.

While developing the general theory of relativity, Einstein's primary goal was to generalize the theory of relativity in such a way as to ensure the equality of all reference frames in the description of nature, regardless of their motion states \cite{Einstein1924}. Of course, general relativity has a second important goal, which is to harmonize the Newtonian theory of gravitation with the theory of relativity. The equivalence principle provided a strong clue that these two goals could be achieved together. In later years, however, the first goal of general relativity received less attention and was less studied. In our opinion, if we judge general relativity considering Einstein's first goal, there are some gaps in the path from the study of accelerated motion to general relativity. Closing these gaps requires knowing the details of acceleration processes and is only possible with a dynamical approach to relativity. The kinematic approach to special relativity deals with inertial frames and the transformations between them, but ignores the previous acceleration processes of frames. However, the details of the acceleration process are essential to understand what happens to the moving rods and clocks. The kinematic approach leaves some questions unanswered. For example, consider two inertial frames that are at rest with respect to each other. Then one of the frames accelerates and come to rest in another inertial frame. Let us denote the inertial frames before and after acceleration as $S_0$ and $S_1$ respectively. Relative to $S_0$, the rods and clocks on $S_1$ have changed with respect to their pre-acceleration state. The clocks have become asynchronous and the rods have contracted. But what physical effects cause clocks and rods to change? For example, the clocks can be modeled with rotating wheels so that one period of the wheel represents a unit of time. Then, where is the torque that changes the rotational phase of the wheels? Our dynamical model provides answers to these questions by considering the details of the acceleration process. Using our dynamical model, we have shown that the equivalence principle and gravitational redshift is necessary to explain special relativistic clock desynchronization. This result is interesting and shows that special relativity by itself is an incomplete theory. Einstein knew that special relativity was incomplete and that's why he tried to generalize his theory. 
Our dynamical model has contributed to filling some of the gaps in the path from the study of accelerated frames in special relativity to general relativity.





\appendix
\section*{Appendix A: Derivation of formulas (\ref{deltalenght}) and (\ref{timeshift}) for infinitesimal proper distance}

Let's consider the clocks A and B as the left and right ends of a rigid rod of proper length $\delta\ell$, where $\delta\ell$ is infinitesimal. From the equations (\ref{xversust}) and (\ref{vversust}) it can be shown that the velocity difference at points A and B is infinitesimal:
\begin{eqnarray}
 v_A-v_B=\delta v=\frac{g^2\;\delta\ell\;t}{c^2 \left(1+\frac{g^2t^2}{c^2}\right)^{3/2}}.
\end{eqnarray}
Since the difference of $\gamma$ factors at A and B points is also infinitesimal, using $\gamma_A$ or $\gamma_B$ in the calculation of length contraction makes no difference. Using equations (\ref{xversust}) and (\ref{vversust}) we get 
\begin{eqnarray}
 \delta x=x_B-x_A=\frac{\delta\ell}{\sqrt{1+\frac{g^2t^2}{c^2}}}=\frac{\delta\ell}{\gamma_A}
\end{eqnarray}
where
\begin{eqnarray}
 \gamma_A=\frac{1}{\sqrt{1-\frac{v_A^2}{c^2}}}=\sqrt{1+\frac{g^2t^2}{c^2}}.
\end{eqnarray}
Thus, the special relativistic length contraction formula is verified.

Similarly, let's calculate the time shift between clocks A and B, observed by $S_0$ at time $t=t_{0A}=\frac{c}{g}\beta_0 \gamma_0$. According to an observer at $S_0$ , the total proper time clock A ticked from $t=0$ to $t=t_{0A}$ is given by eqn.(\ref{tau0A}). On the other hand, the total proper time clock B ticked from $t=0$ to $t=t_{0A}$ is given by
\begin{eqnarray}
 \tau_{B}=\int_0^{t_{0A}} \sqrt{1-\frac{v_B^2}{c^2}} dt_B=\left(\frac{c}{g}+\frac{\delta\ell}{c}\right)\ln \left[\frac{c^2\beta_0\gamma_0+\sqrt{(c^2+g\delta\ell)^2+c^4\beta_0^2\gamma_0^2}}{c^2+g\delta\ell}\right].
\end{eqnarray}
For the infinitesimal $\delta\ell$ the above expression reduces to
\begin{eqnarray}
 \tau_{B}=\left(\frac{c}{g}+\frac{\delta\ell}{c}\right)\left\{\frac{1}{2}\ln \left[\frac{1+\beta_0}{1-\beta_0}\right]-\frac{g\delta\ell \beta_0}{c^2}\right\}
\end{eqnarray}
which gives
\begin{eqnarray}
 \Delta\tau=\tau_B-\tau_A=\frac{\delta\ell}{2c}\ln \left[\frac{1+\beta_0}{1-\beta_0}\right]-\frac{\delta\ell \beta_0}{c}.
\end{eqnarray}
Thus eqn.(\ref{timeshift}) is obtained for infinitesimal $\delta \ell$.

\newpage

\end{document}